\documentclass[nofootinbib,preprint]{revtex4-1}

\usepackage{amsmath}
\usepackage{amsthm}
\usepackage[colorlinks=true,urlcolor=blue,citecolor=blue,linkcolor=blue]{hyperref}
\usepackage[shortlabels]{enumitem}
\usepackage{amsfonts}
\usepackage[final]{graphicx}
\usepackage[caption = true]{subfig}
\usepackage{multirow}
\usepackage{float}
\usepackage{subcaption}
\usepackage{bbold}
\usepackage{stmaryrd}
\usepackage{color}
\usepackage{verbatim}
\usepackage{cases}
\usepackage{amsfonts}
\usepackage{amssymb}
\usepackage[table]{xcolor}
\usepackage{epsfig}
\usepackage{bm}
\usepackage{setspace}
\usepackage{enumerate}
\usepackage{dcolumn}
\usepackage{bm}
\usepackage{enumitem}
\usepackage{multirow}
\usepackage{multirow, array} 

\usepackage{subcaption}

\usepackage{dcolumn}
\usepackage{bm}
\usepackage{amssymb}
\usepackage{lipsum}
\usepackage{amsmath}
\usepackage{graphicx}
\usepackage{array}
\usepackage{multirow}
\usepackage{booktabs}

\usepackage[english]{babel}
\newcommand{\pa}{\partial}

\newcommand{\De}{\Delta}

\def\be {\begin {equation}}
\def\ee {\end {equation}}

\newcommand{\ba}{\begin{array}}
\newcommand{\ea}{\end{array}}
\newcommand{\bea}{\begin{eqnarray}}
\newcommand{\eea}{\end{eqnarray}}
\newcommand{\bi}{\begin{itemize}}
\newcommand{\ei}{\end{itemize}}

\begin{document}

\title{Cylindrically confined hydrogen atom in magnetic field: variational cut-off factor}

\author{A. N. Mendoza Tavera}%
\email{anmtavera@gmail.com}
\affiliation{Departamento de F\'{i}sica, Universidad Aut\'onoma Metropolitana Unidad Iztapalapa, San Rafael Atlixco 186, 09340 Cd. Mx., M\'exico}

\author{H. Olivares-Pil\'on}%
\email{horop@xanum.uam.mx}
\affiliation{Departamento de F\'{i}sica, Universidad Aut\'onoma Metropolitana Unidad Iztapalapa, San Rafael Atlixco 186, 09340 Cd. Mx., M\'exico}

\author{M. Rodr\'iguez-Arcos}%
\email{marisol{\_}ra@tec.mx}
\affiliation{
Tecnológico de Monterrey, Escuela de Ingeniería y Ciencias, Calle del Puente 222, Mexico City 14380, Mexico}

\author{A. M. Escobar-Ruiz}%
\email{admau@xanum.uam.mx}
\affiliation{Departamento de F\'{i}sica, Universidad Aut\'onoma Metropolitana Unidad Iztapalapa, San Rafael Atlixco 186, 09340 Cd. Mx., M\'exico}

\date{\today}

\begin{abstract}
In the present study, we consider the hydrogen atom confined within an impenetrable infinite cylindrical cavity of radius $\rho_{0}$ in the presence of a constant magnetic field ${\bf B} = B\,\hat{\bf z}$ oriented along the main cylinder’s axis. In the Born-Oppenheimer approximation, anchoring the nucleus to the geometric center of the cylinder, a physically meaningful 3-parametric trial function is used to determine the ground state energy $E$ of the system. This trial function incorporates the exact symmetries and key limiting behaviors of the problem explicitly. In particular, it does not treat the Coulomb potential nor the magnetic interaction as a \textit{perturbation}.  The novel inclusion of a variational cut-off factor $\big(1 - \big(\frac{\rho}{\rho_0}\big)^\nu\big)$, $\nu \geq 1$, appears to represent a significant improvement compared to the non-variational cut-off factors commonly employed in the literature. The dependence of the total energy $E=E(\rho_0,\,B)$ and the binding energy $E_b=E_b(\rho_0,\,B)$ on the cavity radius $\rho_0 \in [0.8,\,5] \,$a.u. and the magnetic field strength $B\in [0.0,\,1.0]\,$a.u. is presented in detail. The expectation values $\langle \rho \rangle$ and $\langle|z| \rangle$, and the Shannon entropy in position space are computed to provide additional insights into the system's localization. A brief discussion is provided comparing the 2D and 3D cases as well.
\end{abstract}

\maketitle

\section{Introduction} 

The study of atomic and molecular systems under spatial confined conditions is a growing 
field of research, mainly due to the numerous problems and technological applications 
that can be explored with these models 
{(see for instance~\cite{SBC200957,SBC200958} and references therein)}. 
For instance, atoms and molecules confined 
within nanotube and fullerene traps, zeolitic nanochannels, and the 
behavior of donor centers in quantum-well semiconductor
structures~\cite{ZC1994,YESHC2005,CJ1991,SRS2006,SBC200957,SBC200958}. 
In general, confinement can significantly alter electronic, optical, and magnetic properties, 
leading to phenomena such as shifts in energy levels, changes in reactivity, 
and unique spectroscopic signatures.

The hydrogen atom, being the simplest atomic system with one proton and one electron, 
has been studied considering a wide variety of confining cavities. 
The profile of the corresponding confining potentials includes the
simple harmonic oscillator potential~\cite{PATIL20091},
conoidal boundaries~\cite{LEYKOO200979},
spherical, spheroidal and cylindrical cavities~\cite{SBC200957,LKC:1981,MLGLS:2024},
and Gaussian functions~\cite{Nascimento_2011,OEQA_2023}, among others.

{On the other hand, the influence of a constant magnetic field on the energy spectrum and other physical properties of atomic and molecular systems has been extensively studied~\cite{WB:1999,GD:2006,HGHF:2011}. Similar to spatial confinement, the hydrogen atom has played a central role in these studies~\cite{HGHF:2011,TAH:2009,Thirumalai2025}, serving as a fundamental model for understanding the underlying physics. This naturally leads to the question: how is a physical system affected when it is simultaneously subjected to spatial confinement and exposed to a constant magnetic field?}

{With the simple confinement models mentioned above}, it is also possible to address
the aforementioned systems - nanotube, fullerene traps and zeolitic nanochannels - 
when they are placed in the presence of an external magnetic field $B$. 
The study of systems under spatial confinement conditions in the 
presence of a magnetic field has naturally been explored. For example, we can mention
($i$) the two-dimensional (2D) harmonic oscillator in an electric 
and magnetic field~\cite{ARORA2023},  ($ii$) the 2D hydrogen atom confined to a circular region in the presence of a magnetic 
field $B$ perpendicular to the plane of the electronic motion with
the proton fixed at the center of the circle~\cite{KSSV2020}, ($iii$) hydrogenic impurities in quantum wires in the presence of a magnetic field \cite{NICULESCU2001319, Xiao, PhysRevB.47.1316} (and references there in). To the best of the authors' knowledge, all previous variational studies on impurities in quantum wires have effectively treated the Coulomb interaction as a \textit{perturbation} \cite{NICULESCU2001319, Xiao, PhysRevB.47.1316}. Specifically, the cut-off factor, which ensures that the function satisfies the boundary condition on the cylindrical surface and often expressed through confluent hypergeometric functions, remains unchanged regardless of whether the Coulomb interaction is present or absent. Additionally, the number of variational parameters used in these studies is typically very limited, usually involving only a single variational parameter. In the present approach, we promote the cut-off factor to a variational term, which is then determined by minimizing the energy.

The goal of the present work is to study the 3D
hydrogen atom confined by an impenetrable cylindrical cavity of radius $\rho_0$ 
in the presence of a constant and uniform magnetic field $B$. The magnetic field is parallel to the 
main axis of the cylinder, and the infinitely massive proton is located in the corresponding geometric center. It is worth mentioning that a carbon nanotube serves as a quasi-1D system that confines the hydrogen atom. Hence, understanding electron-proton interactions in spatial cylindrical and magnetic confinements can aid in designing quantum devices using nanotubes as well.

Specifically, for the ground state level, we use the variational method with a simple, physically meaningful 3-parametric trial function to obtain the energy $E=E(B,\rho_0)$. Our aim is not to calculate benchmark results but to achieve reasonably accurate energy estimates (in the non-relativistic domain) using a compact trial function. The uncertainty regarding the complexity of the exact cut-off factor is treated on equal footing with the other variational terms in the trial function, i.e., the cut-off factor ensuring compliance with the boundary conditions on the cylinder's surface is also determined variationally. This natural approach represents a notable distinction from other variational calculations commonly found in the literature on confined systems. For a closely related approach, we refer the reader to \cite{RSR}, where the minimization process in the corresponding cut-off factor was not implemented. It will be shown that this approach significantly enhances previous results reported in the literature. 

{Within the non-relativistic domain and the Born-Oppenheimer approximation (accurate to three significant digits), our proposed ground-state trial function achieves energy values comparable to those obtained from more precise numerical methods, particularly for small magnetic fields and moderate confining radii, with agreement less than $1\%$. The trial function strongly satisfies the cusp condition (the relative deviation is around $10\%$ at $\rho_0 = 2\,a.u.$ and decreases smoothly to less than $5\%$ at $\rho_0 = 5\,a.u.$ across all magnetic field values), remains compact and analytically tractable, and allows for a closed-form interpolation across the parameter space. Notably, we demonstrate for the first time that a variational treatment of the cutoff factor can substantially enhance the accuracy of variational calculations, revealing an important and previously underexplored degree of freedom in trial function design.}

The paper is organized as follows. In section \ref{General}, a three-dimensional  Schrödinger operator $\hat{H}$ is introduced with a generic potential $V(r,\rho)$ which depends only on the two variables $r=\sqrt{x^2+y^2+z^2}$ and $\rho=\sqrt{x^2+y^2}$. The cylindrically confined hydrogen atom in a constant magnetic field is governed by a particular case of this operator. Using the azimuthal symmetry of the system, the original three-dimensional eigenvalue problem for $\hat{H}$ is then reduced to a two-dimensional one. In this section, the so called $(N,m,p)-$representation is defined. The well-known exact solutions of the free ($B=0$), unconfined hydrogen atom are presented in Appendix \ref{nmprep} using these quantum numbers $(N,m,p)$ and compared to the more commonly used representation $(n,m,\ell)$ intrinsic to spherical coordinates. Subsequently, in section \ref{conhyB}, we analyze the Schrödinger operator for a hydrogen atom confined within a cylindrical region under a uniform axial magnetic field $B$, where the vector potential is expressed in the symmetric gauge. Such a gauge naturally occurs in the $(N,m,p)-$representation. For the ground state $(N=0,m=0,p=0)$ the variational energy using a 3-parametric trial function is computed for the cavity radius $\rho_0 \in [0.8,\,5] \,a.u.$ and the magnetic field strength $B\in [0,\,1]\, a.u.\,$. The localization of the electronic cloud is analyzed through the expectation values $\langle \rho \rangle$ and $\langle |z| \rangle$. Finally, in section \ref{shanent}, again just as for the ground state $(N=0,m=0,p=0)$ the Shannon entropy in the position space is calculated.

Throughout the paper atomic units are used $\hbar=m_e=-e=1$.

\section{Generalities}
\label{General}
Let us consider the three-dimensional Hamiltonian operator in Cartesian coordinates $(x,y,z)$
\begin{equation}
\label{Hamiltonian0}
   \hat{H} \ = \ -\frac{1}{2}\,\Delta^{(3)} \
   + \ V\ ,\qquad   \De^{(3)} \ =\ \pa_x^2\ +\ \pa_y^2\ +\ \pa_z^2\ ,
\end{equation}
where the potential $V$ depends on the  variables $r$ and $\rho$ only, 
\begin{equation}
\label{potential}
  V\ = \ V(r,\,\rho)\qquad ; \qquad  r\ =\ \sqrt{x^2+y^2+z^2}\qquad , \qquad \rho=\sqrt{x^2+y^2}\ .
\end{equation}
The  Schr\"odinger equation associated with the Hamiltonian (\ref{Hamiltonian0}) is of the form
\begin{equation}
 \label{schroedinger}
 \hat{H}\,\psi\ =\  E\,\psi \ ,
\end{equation}
where the energy $E$ defines the spectral parameter. {Since the $z$-component of the angular momentum operator, $\hat{L}_z = -i(x\,\partial_y - y\,\partial_x)$, and the $z$-parity operator, defined by ${\hat\Pi}_z\, \psi(x,y,z) = \psi(x,y,-z)$, both commute with the Hamiltonian $\hat{H}$ (\ref{Hamiltonian0}), the eigenfunctions $\psi$ can be written in the form~\cite{JTA2021}
\begin{equation}
\psi(\rho,\,r,\,\varphi) \ = \ \rho^{|m|}\, z^p\, e^{i \,m\, \varphi}\, \Psi(\rho,\,r) \ ,
\end{equation}
where $\Psi(\rho, r)$ is a $\varphi-$independent function to be determined, $\varphi = \tan^{-1}(y/x)$. Substituting this ansatz into the Schrödinger equation (\ref{schroedinger}) yields the reduced eigenvalue problem
\begin{equation}
\label{schroedingerGam}
\mathcal{H}_\Gamma(r,\,\rho,\,\partial_r,\,\partial\rho)\, \Psi(r,\,\rho)\  = \  E\,\Psi(r,\,\rho)\ .
\end{equation}

In terms of the coordinates $(\rho,\,r\,)$, the operator $\mathcal{H}_\Gamma$ reads \cite{JTA2021}
\begin{equation}
\label{schroedinger_new}
\mathcal{H}_\Gamma \, = \,-\frac{1}{2}\left[\partial_\rho^2 + \frac{2\,\rho}{r} \partial^2_{\rho\, r} + \partial_r^2 + \frac{2|m| + 1}{\rho} \partial_\rho + \frac{2(|m| + p + 1)}{r} \partial_r \right]\, +\, V(r,\,\rho) \ ,
\end{equation}
and it explicitly depends on the quantum numbers $m$ and $p$. The corresponding volume element in these coordinates is given by $d^3 \mathbf{r} \propto \frac{r\,\rho}{\sqrt{r^2 - \rho^2}}\, dr\, d\rho\, d\varphi$.

To summarize, each eigenstate $\Psi$ of the two-dimensional operator $\mathcal{H}_\Gamma$ in Eq.~(\ref{schroedingerGam}) is characterized by a triplet of quantum numbers $(N, m, p)$. Here, $m = 0, \pm1, \pm2, \ldots$ is the magnetic quantum number (eigenvalue of $\hat{L}_z$), $p = 0, 1$ corresponds to the eigenvalue $\nu = (-1)^p = \pm 1$ of the $z$-parity operator $\hat{\Pi}_z$, and $N = 0, 1, 2, \ldots$ labels the eigenstates of $\mathcal{H}_\Gamma$. We refer to this classification as the $(N, m, p)$ representation.

Now, for central potentials $V=V(r)$, using spherical coordinates, the natural representation is $(n,\ell,m)$. For clarity, Appendix \ref{nmprep} presents a comparison of both representations in the exactly solvable case of the unconfined free-field Coulomb problem. Furthermore, Appendix \ref{Npmevo} provides a detailed analysis of the evolution of the $(N, m, p)$ representation as a function of the magnetic field and the confinement radius.

As our analysis is solely restricted to the ground electronic state and the electron spin is not explicitly considered, it contributes only as a constant term in the Hamiltonian, we choose to adopt the representation $(N, m, p)$ rather than the conventional spectroscopic notation. This approach omits the spin label, which is unnecessary in this context and would otherwise add redundant complexity.
}

\section{Confined hydrogen atom in a constant magnetic field}
\label{conhyB}

{
In this section we focus on the hydrogen atom (with infinite proton mass) confined by a spherical 
cavity of radius $\rho_0$ in the presence of a magnetic field $B$. In this case, the corresponding potential 
$V(\rho,\,r)$ in (\ref{schroedinger_new}) 
has the following form:
\begin{equation}
\label{Vex}
V(\rho,\,r) \ =  -\frac{1}{r} \ +\  V_{\rm Zeeman} \ + \ V_{\rm conf}\ .
\end{equation}
In the symmetric gauge $\mathbf{A}=\frac{1}{2}\mathbf{B} \times \mathbf{r}$, the second term
\begin{equation}
\label{Vex2}
 V_{\rm Zeeman}  \ =  \frac{m\,B}{2} + \frac{1}{8}\,B^{2}\,\rho^{2}  \ ,
\end{equation}
includes $i$) the linear Zeeman effect $m\,B/2$ due to the interaction of the electron's 
magnetic moment with the magnetic field $B$ (for the ground state where $m=0$, this term 
vanishes) and $ii$) the quadratic Zeeman effect $B^2\rho^2/8$, which is responsible for the increase in energy at strong magnetic fields and dominates the overall energy behavior in this regime. At large $B \rightarrow \infty$, the ground state energy of the unconfined hydrogen 
atom goes as $E=-\frac{1}{2}\log^2{(B/B_0)}$.
In the present study, the magnetic field strength $B$  is expressed in atomic units, where $B_0 \approx 2.35 \times 10^5\,\text{T} = 2.35 \times 10^9\,\text{G}$. For example, in Eq.~(\ref{Vex2}), $B$ is given in units of $B_0$, and \( m \) denotes the magnetic quantum number.

It is important to note that, unlike the Coulomb potential, the effective potential $ V_{\rm Zeeman}$ is gauge-dependent. Consequently, a change in the gauge leads to a different expression for $ V_{\rm Zeeman}$. Nevertheless, physical observables such as the energy, which are gauge-independent, remain unaffected. The results obtained in different gauges are related through well-known unitary transformations, ensuring the consistency of the physical predictions.

}

\vspace{0.1cm}

\begin{figure}
    \includegraphics[width=0.8\linewidth]{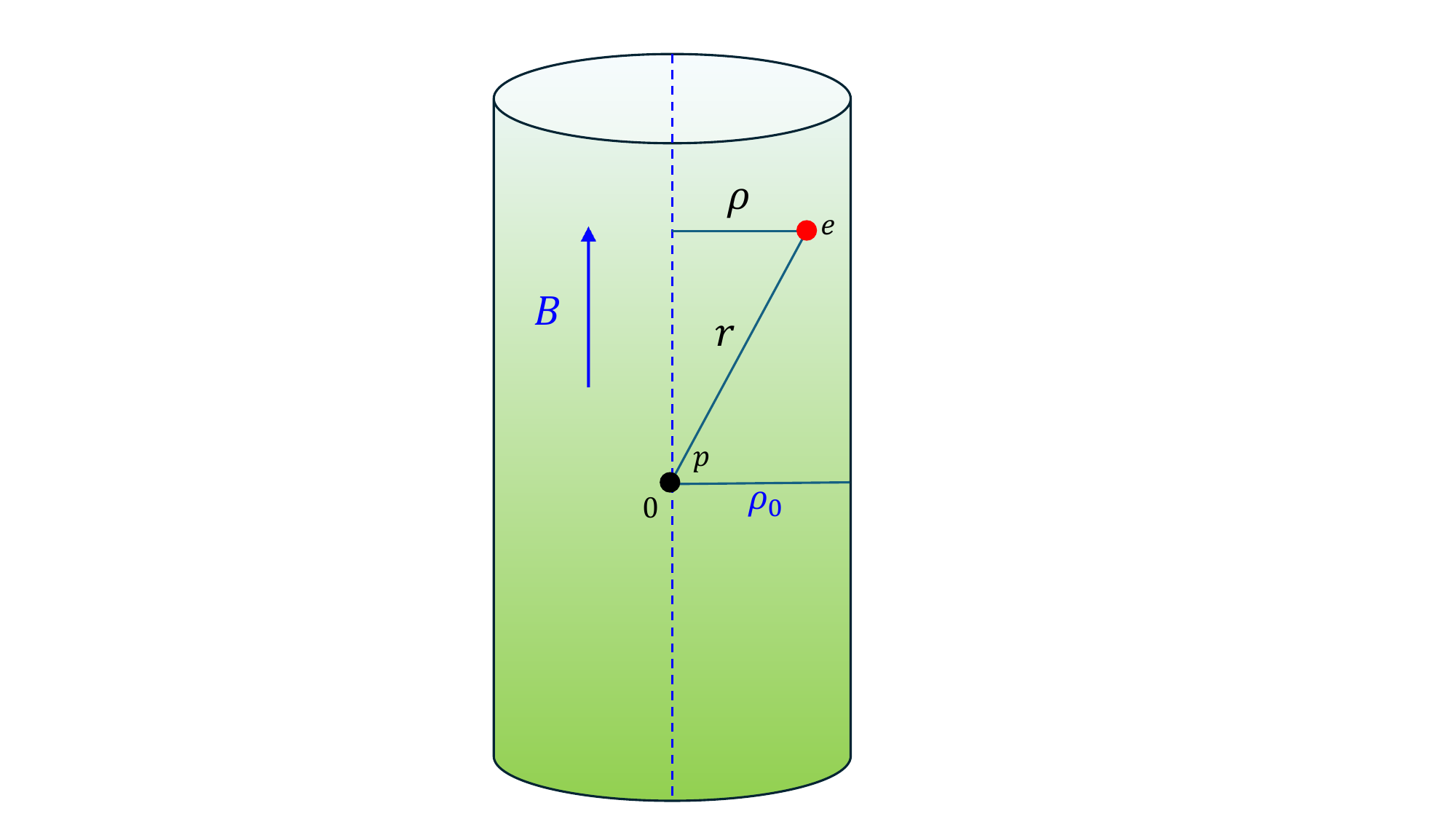}
    \caption{\small Geometric configuration of the cylindrically confined hydrogen atom in a constant magnetic field. The proton ($p$) is located at the origin, and the magnetic field is aligned along the cylinder's axis. The variables $\rho$ and $r$, for the electron ($e$), that naturally appear in the potential (\ref{Vex}) are displayed as well. The radius of the infinite cylinder is denoted by $\rho_0$.}
    \label{GeoSet}
\end{figure}

Additionally, we assume that the system is confined in an infinite cylindrical impenetrable shell of radius $\rho_0$. This is modeled by the confining potential
\begin{equation}
    V_{\rm conf} \  = \  
\left\{
\begin{array}{ll}
0 & \text{if } \rho < \rho_0\ , \\
\infty & \text{if } \rho \geq \rho_0 \ .
\end{array}
\right.
\end{equation}
The longitudinal axis of the cylinder as well as the magnetic field are directed along the $z-$direction. The proton is located at the origin, see Fig. \ref{GeoSet} for the geometrical settings. This is the potential (\ref{Vex}) we will consider in detail. It depends on two external quantities $B$ and $\rho_0$. To solve the associated Schrödinger equation, we employ the variational method (see below).

\subsection{Trial function}

To perform concrete variational calculations we use the two-dimensional Hamiltonian $ {\cal H}_\Gamma$ (\ref{schroedinger_new}) with potential (\ref{Vex}). The state with positive parity $(m=0,p=0)$, the ground state, is studied in detail only. To this end, we employ the following trial function:
\begin{equation}
\label{psivar}
\Psi(\rho,r)\ = \ F(\rho)\ e^{-\alpha\,r-\beta\,B \,\rho^2}\ ,
\end{equation}
which contains the product of Coulomb- and Laudau-like orbitals, $e^{-\alpha\,r}$ and $e^{-\beta\,B \,\rho^2}$, respectively. The cut-off factor in (\ref{psivar})
\begin{equation}
\label{ctF}
F(\rho) \ = \ \bigg[1\,-\,{\bigg(\,\frac{\rho}{\rho_0}\bigg)}^{\nu}\,\bigg] \ ,
\end{equation}
accounts for the boundary condition on the surface of the cylinder, namely $\Psi(\rho=\rho_0,\,r)=0$. In (\ref{psivar}), $\alpha,\,\beta,\,\nu$ are variational parameters. At zero magnetic field ($B=0$), we recover the exact solution $e^{-r}$ for the free unconfined system ($\rho_0 \rightarrow \infty$, thus, $F \rightarrow 1$) with values $\beta=0$ and $\alpha=1$. Likewise, the exact solution for the limiting case \( \rho_0 \to \infty \) and vanishing Coulomb interaction is recovered when \( \beta = \frac{1}{4} \). For the original three-dimensional Hamiltonian (\ref{Hamiltonian0}) the complete trial function takes the form
\begin{equation}
\label{psit}
\psi(\rho,r,\varphi)  \ = \ \rho^{|m|}\,z^p\, e^{i\,m\,\varphi}\,\bigg[1\,-\,{\bigg(\,\frac{\rho}{\rho_0}\bigg)}^{\nu}\,\bigg]\,e^{-\alpha\,r-\beta\,B \,\rho^2} \ .
\end{equation}

{
Although a factor ${\big[1\,-\,\frac{\rho}{\rho_0}\,\big]}^{\nu}$ seems a reasonable choice and was initially tested, it neither improves the variational energy nor accurately preserves the cusp condition; in contrast, our proposed cutoff factor in (\ref{psit}) not only yields better energy results but also more accurately preserves the cusp condition and more smoothly recovers the free case behavior as it approaches unity in the unconfined free field limit.

}

\textbf{Remark.}\textit{\ Here, unlike conventional studies in the literature, our aim is to identify the optimal choice within an infinite one-parametric family of cut-off factors. We do not select it rigidly; it arises naturally as a result of minimizing the energy. To the best of the authors' knowledge, a genuine variational cut-off factor in confined atomic physics is introduced for the first time. It will be shown that this approach significantly enhances the accuracy of variational results.} 

It is worth noting that the standard Coulomb-like and Landau-like ground state orbitals arise from the asymptotic analysis of the exact solutions for the unconfined Coulomb and Landau problems, respectively, at large distances. However, because the system under study is spatially confined within an impenetrable cylindrical cavity, these orbitals are no longer uniquely determined by the normalizability conditions. This important conceptual aspect is rarely discussed in the literature. In Section \ref{sizeatom}, a quantitative analysis is performed to determine the limits of reasonable physical applicability of (\ref{psit}).


Building upon the previous argument,
a comparison of the variational calculations for the ground state 
($m=0,p=0$) at zero magnetic field $B=0$ is presented in Table~\ref{Tcomp}
for the confining radii $\rho_0=2.0$ and $\rho_0=4.0$~a.u..
Traditionally the value of the parameter $\nu$ in the trial function (\ref{psivar}) 
is taken as $\nu=1$. However, it can be observed that considering different 
values of $\nu$ greater than 1 ($\nu > 1$), 
a significant improvement in the variational energy $E_{var}$ is found.
In order to obtain a better value of the variational energy, the constant $\nu$ is elevated to the status of an additional variational parameter in the trial function (\ref{psivar}). Note that the traditional cases $\nu=2,3$ also do not yield the optimal energy values. The results of the variational energy (and the optimal values of the parameters $\alpha$ and $\nu$) 
are presented in bold in Table~\ref{Tcomp}.
{
The same system, the hydrogen atom in a cylindrical potential, was considered 
in~\cite{Ndengue2014} by B-Spline based methods, showing an agreement of at least 
three significant figures with the calculations previously reported in~\cite{YSP:2008}
where the finite-difference method and the collocation method were employed.
However, in the particular case of a cylindrical cavity of infinite height 
a comparison with the present results is possible only with those reported 
in~\cite{Ndengue2014}, giving an agreement of $\sim 0.9$\% and $\sim 0.03$\% 
for radii $\rho=2.0$ a.u. and $\rho=4.0$ a.u., respectively.}

\begin{table}[H]
\begin{center}
\begin{tabular}{|l l c|l l c| }
\hline
\multicolumn{3}{|c}{$\rho_{0}=2.0\,$[a.u.]} \vline &
\multicolumn{3}{c|}{$\rho_{0}=4.0\,$[a.u.]} \\
\hline\hline
$E_{var}$& $\nu$& $\alpha$&$E_{var}$& $\nu$& $\alpha$\\
\hline
-0.1745  & 1     &0.7497&-0.4654  &1     &0.7835\\
 -0.2613  & 2     &0.9387&-0.4853  &2     &0.9095\\
-0.2757  & 3     &1.0612&-0.4894  &3     &0.9640\\
{\bf -0.2767}&{\bf 3.4951}&{\bf 1.1058}&{\bf -0.4917}&{\bf 6.1199}&{\bf 1.0142}\\
\hline
-0.279120&\cite{Ndengue2014}&&-0.491863&\cite{Ndengue2014}&\\
\hline
\end{tabular}
\label{tab:energiascompa0}
\end{center}
\caption{
Ground state energies $(m=0,p=0)$  for  the confinement radii 
$\rho_{0} = 2.0$ and $4.0\,$a.u. at zero magnetic field $B = 0$. A comparison of the 
energies corresponding to fixed values of $\nu = 1, 2, 3$ in the cut-off function is shown. 
When $\nu$ parameter is optimized, the results are shown in bold. 
Results reported in~\cite{Ndengue2014} are displayed as well. {Positive-energy states lie in the continuum and are interpreted as resonances, rather than genuine bound states.}}
\label{Tcomp}
\end{table}


\subsection{Zero magnetic field $B=0$}

\begin{center}
\begin{figure}
\subfloat{\includegraphics[width=0.55\linewidth]{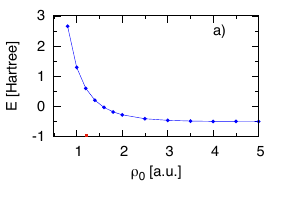}} \\
\subfloat{\includegraphics[width=0.55\linewidth]{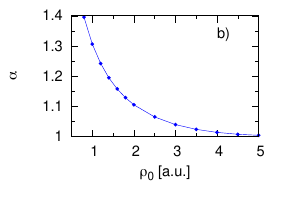}}
\subfloat{\includegraphics[width=0.55\linewidth]{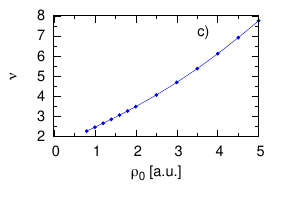}}
\caption{{
$a$) Variational energy $E$ for the ground state ($m=0$, $p=0$) of the hydrogen 
    atom confined by an impenetrable cylinder of radius $\rho_0$ at zero 
    magnetic field $B=0$. The small red square correspond to the expectation 
    value for the free case $\langle \rho \rangle_{\rho_{0}=\infty}$.
    The optimal values of the parameters $\alpha$ and $\nu$ are depicted in
    $b$) and $c$), respectively. Positive-energy states lie in the continuum and are interpreted as resonances, rather than genuine bound states.}}
\label{B0}
\end{figure}
\end{center}

Let us first consider the results of the hydrogen atom confined by an infinite 
cylinder of radius $\rho_0$ at zero magnetic field $B=0$. In this case, the 
variational parameter $\beta$ is identically zero, $\beta=0$. In 
Table~\ref{tab:param_m0_p0_B0}, the values of the total energy $E$ and the optimal 
values of the parameters $\alpha$ and $\nu$ are displayed as a function of the 
confining radius $\rho_0 \in [0.8, 5.0]$\,a.u.. 
{
Two limiting cases can be highlighted: $i$) For small confinement radii the spectrum 
is mainly determined by the presence of the cavity and the Coulomb interaction can be 
considered as a perturbation. In  this domain, similar to the 
problem of a particle in an infinite box, the energy value increases as the cavity radius 
decreases. 
$ii$) On the other hand, for large confinement radii the spectrum is mainly determined by the Coulomb interaction and the presence of the cavity slightly modifies the spectrum, see also the discussion presented at the end of this subsection.
This is precisely the behavior that is observed}
in Table~\ref{tab:param_m0_p0_B0}, the total energy $E$ decreases 
monotonically from $E=2.658$~a.u. 
to $E= -0.498$~a.u. as the confining radius increases from $\rho_0=0.8$~a.u.
to $\rho_0 =5.0$~a.u. (see also Fig. \ref{B0}), asymptotically approaching the 
value of the free hydrogen atom energy ($B=0$ and $\rho_0 \rightarrow\infty$),
$E=-1/2$~a.u..

{
Explicitly, for \( B = 0 \) and \( \rho_0 = 21 \, \text{a.u.} \), direct energy minimization yields optimal parameters \( \alpha = 1.0000 \) and \( \nu \approx 16 \), corresponding to an energy of \( E = -\frac{1}{2} \, \text{a.u.} \), the ground-state energy of the free hydrogen atom in the \( 1^{2}S_{1/2} \) state (atomic spectroscopy notation). This alignment allows for a clearer and more intuitive understanding of the system's quantum behavior in this limiting regime.
}

As a function of the confining radius $\rho_0$, the optimal value of the parameter 
$\alpha=\alpha(\rho_0)$, has a smooth behavior, approaching asymptotically from above 
to the limiting value corresponding to the free hydrogen atom $\alpha=1$ as the 
confining radius $\rho_0$ increases (see Fig.~\ref{B0}).
The fulfillment of the cusp condition, governed by $\alpha$, deteriorates as the 
radius $\rho_0$ decreases. However, even at $\rho_0=2.0$\,a.u. the relative error 
with respect to the exact value for the electron density $n(r)$
\begin{equation}
Z=-\frac{1}{2 n(r)}\frac{d n(r)}{dr}\Bigr|_{r\rightarrow 0}    
\end{equation}
$Z=1$ is only of $10\%$. 

Remarkably, the value of the optimal parameter $\nu$ which dictates the functional 
form of the cut-off factor (\ref{ctF}) increases (almost linearly) from 
$\nu\sim 2.3$ at $\rho_0=0.8$\,a.u. up to $\nu \sim 7.8$ at $\rho_0 = 5.0$\,a.u., 
hence, more than 3 times! (see Fig.~\ref{B0}). More importantly, this parameter 
significantly alters the optimal value of the ground state energy, as shown in 
Table \ref{Tcomp}.

When $B=0$, a rough estimate of the applicable domain of (\ref{psivar}) is described by $\rho_0 > \langle \rho \rangle_{\rho_{0}=\infty}=1.178$\,a.u., see Section \ref{sizeatom}.

\begin{table}[!h]
\begin{center}
\begin{tabular}{| c | c| c | c |}
\hline
\multicolumn{4}{|c|}{$B=0$} \\
\hline
$\rho_{0}$~[a.u.] & $E$~[a.u.] & $\alpha$& $\nu$ \\ 
\hline\hline
0.8& \,\,2.658& 1.396& 2.274\\
1.0& \,\,1.295& 1.307& 2.469\\ 
1.2& \,\,0.599& 1.243& 2.666\\ 
1.4& \,\,0.207& 1.195& 2.865\\
1.6& -0.029& 1.158& 3.069\\
1.8& -0.179& 1.129& 3.278\\ 
2.0& -0.277& 1.106& 3.495\\ 
2.5& -0.405& 1.065& 4.070\\ 
3.0& -0.458& 1.040& 4.698\\
3.5& -0.481& 1.024& 5.380\\ 
4.0& -0.492& 1.014& 6.120\\ 
4.5& -0.496& 1.008& 6.916\\
5.0& -0.498& 1.004& 7.764\\  
\hline
\end{tabular}
\caption{Variational energy of the ground state ($m=0$, $p=0$) of the hydrogen 
atom as a function of the cylinder radius $\rho_0$ at zero magnetic field 
$B = 0$. The values of the optimal parameters $\{\alpha,\,\nu\}$ are also shown. {Positive-energy states lie in the continuum and are interpreted as resonances, rather than genuine bound states.}}
\label{tab:param_m0_p0_B0}
\end{center}
\end{table}
{
In the asymptotic regime $\rho_0 \gtrsim 2.5$, where the effect of the cylindrical confinement becomes progressively weaker, the ground state energy $E(\rho_0)$ tends toward the well-known value of the unconfined hydrogen atom, $E_\infty = -0.5$ Hartree. To quantify the finite-size corrections due to the transverse boundary, we performed a fit to the form
\[
E(\rho_0) \approx -0.5 + \frac{A}{\rho_0^\alpha},
\]
obtaining $A \approx 0.4$ and $\alpha \approx 2$. The nearly quadratic decay of the energy correction with $\rho_0$ indicates that the primary confinement effect arises from the restriction imposed in the transverse ($\rho$) direction, while the longitudinal ($z$) motion remains effectively unconfined. This behavior is closely related to the energy scaling of a two-dimensional hydrogen atom subject to Dirichlet boundary conditions (i.e., hard-wall confinement), where the energy levels also exhibit a $1/\rho_0^2$ dependence due to the dimensional reduction of the Coulomb interaction and the confinement-induced quantization of the radial motion, see \cite{PAmore}.

Physically, for large $\rho_0$, the electron is still confined in the transverse plane, but the radial wavefunction extends over a larger region, allowing the system to recover the $1/r$ Coulombic character over most of the accessible space. However, the hard-wall boundary at $\rho = \rho_0$ forces the wavefunction to vanish at the cylinder edge, leading to a perturbative shift in energy that scales inversely with the square of the effective transverse size. This mechanism is analogous to the well-known confinement energy in quantum dots or nanowires with Coulomb potentials.

This energy correction arises because even in the absence of the Coulomb interaction, the wavefunction is forced to vanish at the cylinder wall due to the Dirichlet boundary condition. Thus, it represents the quantization energy from the transverse confinement: it is purely kinetic and dominates the correction to the total energy in the large $\rho_0$ limit, since the longitudinal and Coulomb parts remain essentially unaffected. The resulting solution corresponds to a mode of a two-dimensional cylindrical drum, whose eigenfunctions are Bessel functions of the first kind and whose eigenvalues scale as $\sim 1/\rho_0^2$. This is exactly analogous to the case of a particle in a one-dimensional infinite potential well, where the quantized kinetic energy levels scale as $\sim 1/L^2$, with $L$ being the length of the box.

Therefore, the observed power-law behavior not only confirms the accuracy of the variational approach in the weak confinement regime but also provides a bridge between three-dimensional and lower-dimensional Coulomb systems under geometric constraints. It highlights how the interplay between dimensionality, boundary conditions, and long-range interactions governs the spectral properties of confined atomic systems.

}

\subsection{Magnetic field $B\neq0$}

\begin{figure}
\subfloat[]{\includegraphics[scale = 1.65]{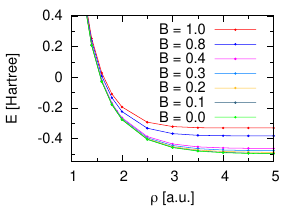}}
\subfloat[]{\includegraphics[scale = 1.65]{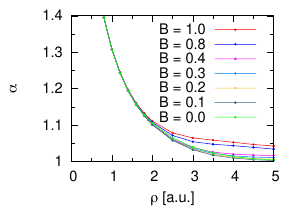}}\\
\subfloat[]{\includegraphics[scale = 1.65]{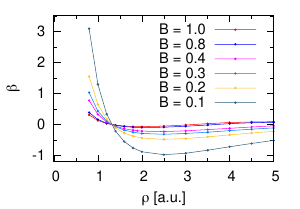}}
\subfloat[]{\includegraphics[scale = 1.65]{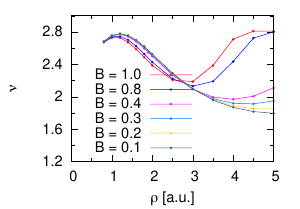}}\\
\subfloat[]{\includegraphics[scale = 1.65]{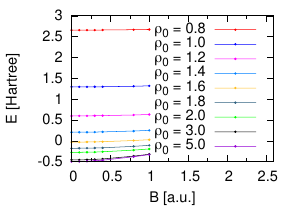}} 
\caption{Variational energy $E$ and optimal parameters 
    $\{\alpha, \beta, \nu\}$ for the ground state ($m=0$, $p=0$) of 
    the hydrogen atom as a function of the cylinder radius $\rho_0$ 
    at different fixed magnetic fields $B$. Figure (e)
    presents the energy as a function of the magnetic field for
    fixed values of the cylindrical cavity radii $\rho_0$. {Positive-energy states lie in the continuum and are interpreted as resonances, rather than genuine bound states.}}
\label{parmsBB}
\end{figure}

Let us now consider the ground state of the cylindrically confined hydrogen 
atom placed in a non-zero magnetic field $B$ parallel to 
the cylinder's axis. Table~\ref{tab:param_m0_p0_Bxx} presents the variational 
energy $E$ as well as the value of the optimal parameters $\{\alpha, \beta, \nu\}$ 
as a function of the confinement radius $\rho_0$ for three values of the 
magnetic field  $B=0.4$, $0.8$ and $1.0$~a.u. (see also Figure \ref{parmsBB}).

\begin{table}[t!]
\begin{center}
\begin{tabular}{| c | c | c | r | c || c | c | r | c || c | c | r | c |}
\hline\hline
&\multicolumn{4}{c||}{$B=0.4$\,[a.u.]}&\multicolumn{4}{c||}{$B=0.8$\,[a.u.]}&\multicolumn{4}{c|}{$B=1.0$\,[a.u.]} \\
\hline\hline
$\rho_{0}$ [a.u.] & $E$ [a.u.] & $\alpha$ & $\beta$\,\,\,\,\,& $\nu$ &
$E$ [a.u.] & $\alpha$ & $\beta$\,\,\,\,\,& $\nu$ &
$E$ [a.u.] & $\alpha$ & $\beta$\,\,\,\,\,& $\nu$ \\
\hline
0.8&  2.659& 1.395&  0.772& 2.682 &  2.667& 1.396&  0.385& 2.677 &  2.673& 1.396&  0.308& 2.673\\
1.0&  1.298& 1.308&  0.321& 2.754 &  1.310& 1.308&  0.159& 2.741 &  1.319& 1.309&  0.126& 2.731\\
1.2&  0.604& 1.244&  0.083& 2.772 &  0.621& 1.245&  0.040& 2.747 &  0.633& 1.245&  0.031& 2.728\\
1.4&  0.214& 1.195& -0.055& 2.745 &  0.236& 1.196& -0.029& 2.703 &  0.253& 1.198& -0.023& 2.675\\
1.6& -0.020& 1.157& -0.138& 2.682 &  0.008& 1.159& -0.068& 2.626 &  0.028& 1.161& -0.053& 2.589\\
1.8& -0.168& 1.127& -0.188& 2.597 & -0.134& 1.130& -0.089& 2.530 & -0.109& 1.133& -0.068& 2.488\\
2.0& -0.263& 1.103& -0.216& 2.501 & -0.223& 1.108& -0.099& 2.429 & -0.194& 1.112& -0.072& 2.388\\
2.5& -0.387& 1.062& -0.230& 2.268 & -0.332& 1.072& -0.090& 2.222 & -0.293& 1.079& -0.055& 2.208\\
3.0& -0.435& 1.038& -0.206& 2.096 & -0.368& 1.055& -0.058& 2.132 & -0.322& 1.065& -0.018& 2.189\\
3.5& -0.453& 1.026& -0.167& 1.998 & -0.378& 1.048& -0.017& 2.192 & -0.329& 1.059&  0.025& 2.388\\
4.0& -0.461& 1.021& -0.125& 1.970 & -0.381& 1.044&  0.024& 2.437 & -0.330& 1.053&  0.061& 2.712\\
4.5& -0.463& 1.018& -0.083& 2.008 & -0.382& 1.039&  0.054& 2.725 & -0.330& 1.047&  0.079& 2.811\\
5.0& -0.464& 1.017& -0.045& 2.111 & -0.382& 1.035&  0.070& 2.802 & -0.330& 1.043&  0.089& 2.810\\
\hline\hline
\end{tabular}
\caption{Variational energy and optimal parameters $\{\alpha, \beta, \nu\}$ for 
the ground state ($m=0$, $p=0$) of the hydrogen atom as a function of the 
radius of the cylindrical cavity $\rho_0$ for magnetic fields $B=0.4, 0.8$ and $1.0$~a.u.; {positive-energy states lie in the continuum and are interpreted as resonances, rather than genuine bound states.}}
\label{tab:param_m0_p0_Bxx}
\end{center}
\end{table}

In the domain considered in this study, in the space $(B, \rho_0)$, 
the ground-state energy $E$ is a smooth function of both, the cavity radius 
$\rho_0$ and the magnetic field $B$. 
For instance, Figures \ref{parmsBB} (a)-(d) depict the dependence of $(E,\alpha,\beta,\nu)$ as a function 
of the cavity radius $\rho_0$ for fixed values of the magnetic field $B$ while 
Figure~\ref{parmsBB} (e) shows the energy $E$ as a function of the 
magnetic field $B$ for different values of the cavity radius $\rho_0$. 
At a fixed value of the magnetic field $B \in[0.0,1.0]$~a.u., the energy $E$ is a monotonically decreasing function of $\rho_0$ (see Figure~\ref{parmsBB} (a)), 
whereas at a fixed radius $\rho_0 \in [0.8,5.0]$\,a.u., the energy $E$ increases 
slightly as the magnetic field strength $B$ grows from $0.0$ to $1.0$\, a.u.\,. 
Indeed, for small values of $\rho_0 \leq 2$\,a.u., the energy $E$ remains almost 
constant as a function of the magnetic field $B$, see Figure~\ref{parmsBB} (e). 

{
For small radii ($\rho_0 \lesssim 2$), the strong transverse confinement raises the energy significantly, and for moderate to large magnetic fields ($B \gtrsim 0.3$), the ground state energy becomes positive. This indicates that the system transitions from a truly bound state to a resonance-like state, where the electron becomes axially delocalized. As $\rho_0$ increases, the confinement weakens and the energy decreases, approaching the value of the unconfined hydrogen atom in a magnetic field. Additionally, for fixed $\rho_0$, increasing $B$ results in higher energy due to magnetic compression of the transverse wavefunction. These results highlight the existence of two distinct physical regimes: a bound-state regime for $\rho_0 \gtrsim 3$, and a resonance regime for $\rho_0 \lesssim 2$, where the variational method—although still providing normalizable trial functions—no longer accurately reflects the physical character of the state.

}

The variational parameter $\alpha$, which measures the screening of the electron 
charge, is a smooth monotonous function of both the cavity radius $\rho_0$ 
(see Figure~\ref{parmsBB} (b)) and the magnetic field $B$. For 
large $\rho_0 \geq 5.0$\,a.u., its optimal value remains close to $1.0$\,a.u., 
and for fixed $\rho_0 \leq 2.0$\,a.u. is almost unaffected by the magnetic 
field $B\in[0.0,1.0]$\,a.u.. Moreover, $\alpha$ displays a behavior similar 
to that of the ground state energy $E$ as a function of the cylinder radius
$\rho_0$ and the magnetic field $B$.

In turn, the parameter $\beta$ exhibits a regular behavior 
(see Figure~\ref{parmsBB} (c)) i.e., no sharp change occurs as a function of 
both the radius of the cylindrical cavity $\rho_0$ and the magnetic field $B$.
In Figure~\ref{parmsBB} (d), fixing the value of the magnetic field $B$, the 
variation of $\nu$ with respect to the radius $\rho_0$ is shown. The optimal 
value in $\nu$ runs from $\nu \approx 1.8$ to $\nu \approx 2.8$ in the domain 
considered. 

\vspace{0.2cm}

\begin{figure}
    \includegraphics[scale = 0.44]{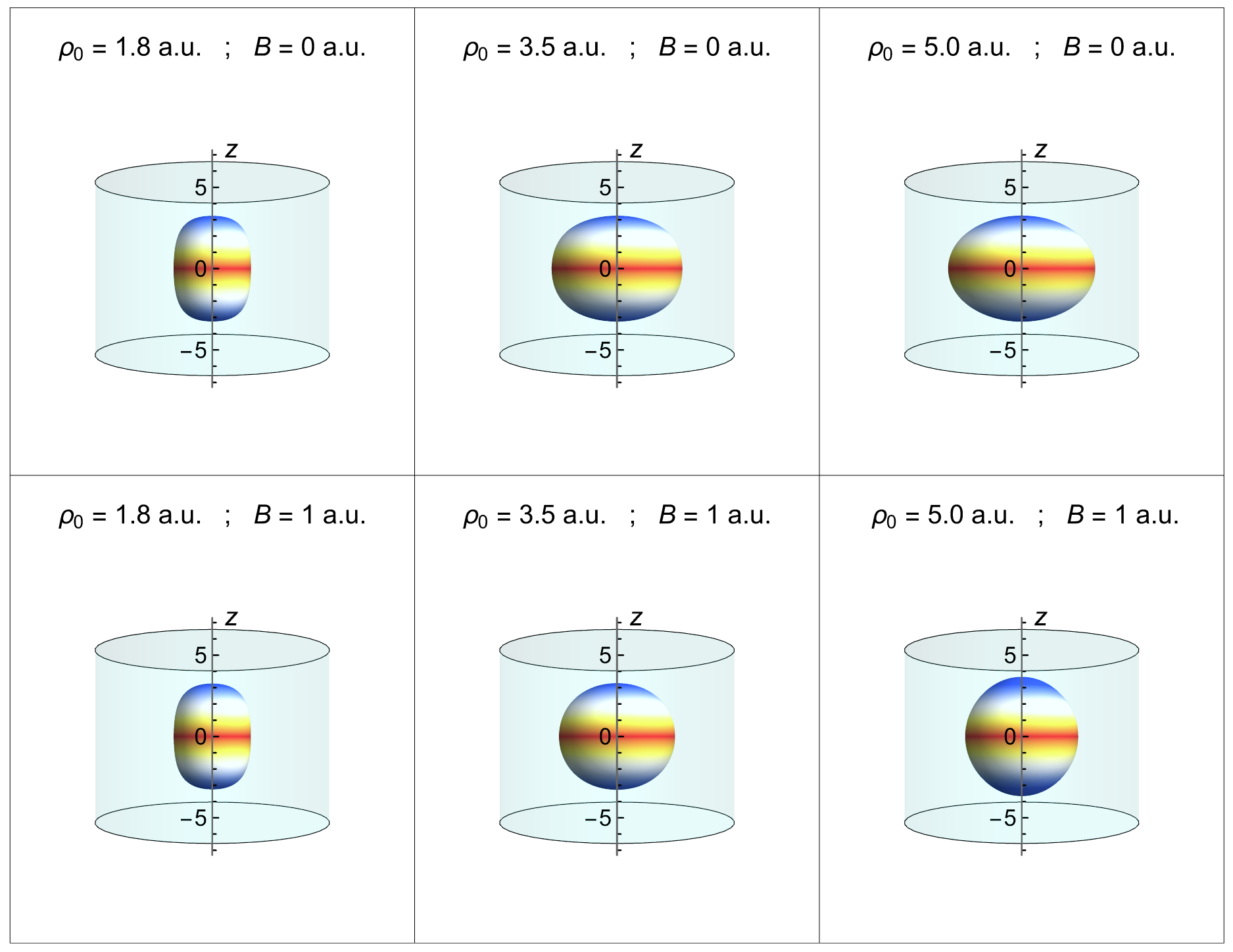}
    \caption{\small Ground state orbital configurations for different magnetic fields and confining radii.}
    \label{orbitalsB}
\end{figure}

{

The behavior of the optimal variational parameters $\alpha$, $\beta$, and $\nu$ as functions of the variational parameter $\rho$ provides insight into the regions where the trial wavefunction is most reliable. As shown in Fig.~\ref{parmsBB}, all three parameters exhibit stable and physically meaningful trends for $\rho \gtrsim 2$~a.u., with minimal fluctuations and smooth dependence on the magnetic field strength $B$. Notably, while some instability is observed at lower $\rho$, particularly in the parameters $\beta$ and $\nu$, this regime does not correspond to the physical scenarios of interest. For truly bound states (see \ref{continuums}) the optimal value $\rho_0$ consistently lies above 2~a.u., ensuring that the wavefunction is being evaluated in its most reliable region. Therefore, the variational ansatz used in this work is well-suited for accurately describing the bound states under consideration.

Ground state orbital configurations for different magnetic fields and confining radii are displayed in Figure \ref{orbitalsB}. Each panel shows the spatial distribution of the ground state wavefunction within a cylindrical potential of radius \( \rho_0 \) (in atomic units, a.u.), with and without an external magnetic field \( B \) (a.u.) applied along the \( z \)-axis. The top row corresponds to \( B = 0 \) a.u., while the bottom row shows results for \( B = 1 \) a.u. Increasing \( \rho_0 \) from left to right (1.8, 3.5, and 5.0 a.u.) allows the orbital to expand radially, reflecting weaker confinement. The presence of a magnetic field tends to localize the electron wavefunction more tightly along the radial direction, counteracting the delocalization caused by weaker confinement.}

By construction, unlike the Landau-like orbital $e^{-\beta\,B\,\rho^2}$, 
the $\nu$-dependent cut-off factor does not correspond to a well-defined physical 
limit. The exact solution in the $\rho-$direction of a particle confined 
to move inside an infinite cylindrical potential is expressed in terms of 
Bessel function of the first kind. Now, a detailed numerical exploration 
reveals that the optimal energy $E$ is highly sensitive to the value of 
$\alpha$, whereas different pairs $(\beta,\nu)$ with the same $\alpha \sim 1$ 
can produce nearly identical energy values. That is, for $\alpha\sim 1$, 
the mixture of spatial and magnetic confinement prevents a clear physical 
interpretation of the parameters $(\beta,\nu)$ separately.    

As described in \cite{NICULESCU2001319}, an alternative trial function 
$\psi_t^{(\rm alt)}$ can be constructed from the exact ground-state 
eigenfunction of the Landau problem within a cylinder (i.e., an electron 
confined by an impenetrable cylindrical boundary in the presence of an 
axial magnetic field) and multiplying it by the factor 
$e^{-\frac{r}{\lambda}}$, where $\lambda$ serves as the sole variational 
parameter. For the cases $\rho_0=1.0,\,1.4,\,2.0$ and $5.0$\,a.u., the 
ground state $(m=0,p=0)$ energy obtained from our trial function (\ref{psivar}) 
shows a relative improvement of $8\%,\,34\%,\,17\%$ and $2\%$, respectively, 
compared to the corresponding values derived from the aforementioned 
$\psi_t^{(\rm alt)}$. It is a clear indication, in spite of its simplicity, 
of the high quality of (\ref{psivar}).

\subsubsection{Binding energy $E_b$}
\begin{figure}
\subfloat[]{\includegraphics[scale = 0.48]{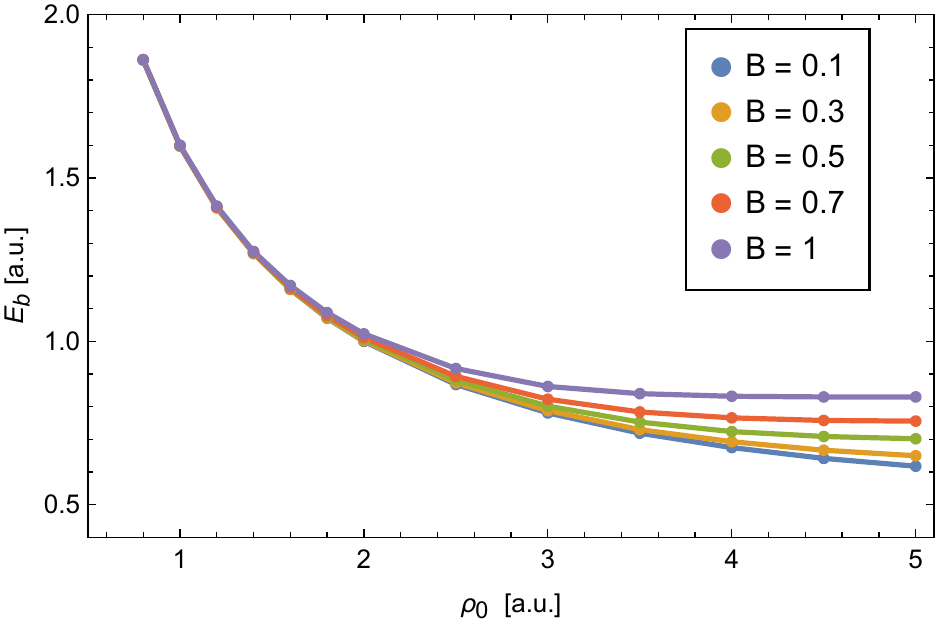}}
\hspace{0.2cm}
\subfloat[]{\includegraphics[scale = 0.48]{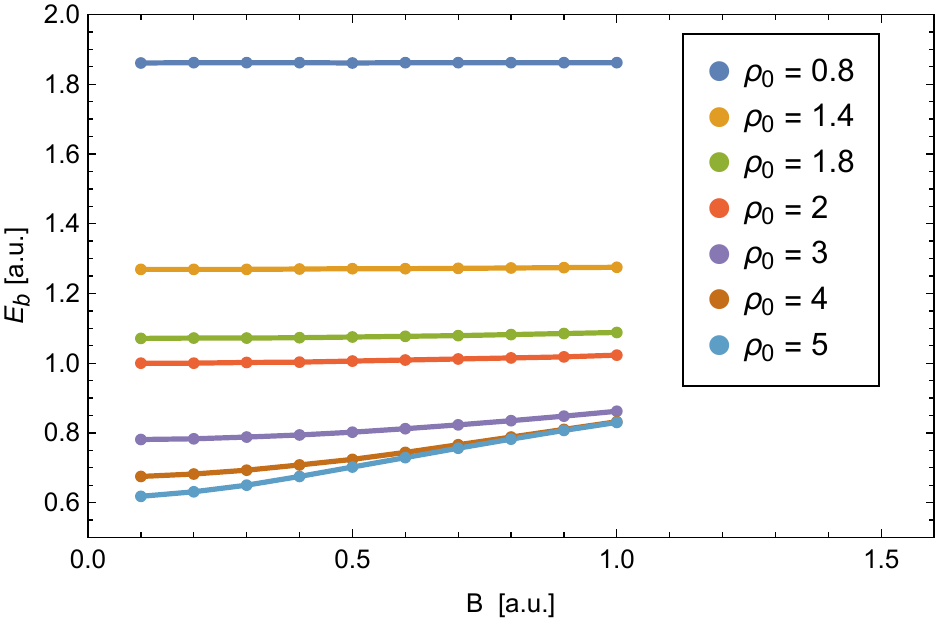}} 
\caption{Binding energy $E_b$ of the confined hydrogen atom in the ground state 
$m=0$, $p=0$ (a) as a function of the radius of the cylindrical cavity $\rho_0$ 
for fixed values of magnetic field $B$ and (b) as a function of the magnetic 
field $B$ for fixed values of the cylindrical cavity radius $\rho_0$.}
\label{Ebb}
\end{figure}

As usual, the binding energy $E_b$
\begin{equation}
E_b  \ = \ E_0 \ - \ E \ ,    
\end{equation}
defined as the difference between the ground state energy $E_0$ of the electron 
in an axial constant magnetic field confined within an impenetrable cylinder 
only and the energy $E$ when the Coulomb interaction is introduced. As is well 
known, for the ground state, $E_0$ can be obtained in terms of the confluent 
hypergeometric function $F(a,b,z)$~\cite{Landau:1991wop} as 
\begin{equation}
F\left[-\left(\frac{E_0}{B}-\frac{1}{2}\right),1,\frac{B}{2}\,\rho_0^2\right]=0\,.
\label{e0hf}
\end{equation}
It is worth mentioning that the absolute difference between the value obtained 
for $E_0$ with the expression~(\ref{e0hf}) and a variational calculation with 
the trial function~(\ref{psit}) is of order $\sim 10^{-4}$.
Figures~\ref{Ebb} (a) and (b) depict the binding energy $E_b$ as a function 
of the cylinder radius $\rho_0$ and the magnetic field $B$, respectively.
As expected, the binding energy $E_b$ decreases monotonically with the increase 
of $\rho$ while it increases gradually with the increase of the magnetic field $B$.

{

\subsubsection{Continuum states}
\label{continuums}

In the presence of a uniform magnetic field aligned $B>0$ along the $z$-axis and an impenetrable cylindrical boundary of radius $\rho_0$, the hydrogen atom exhibits a hybrid confinement regime in which the rotational symmetry of the Coulomb potential is broken by both the field and the geometry. The continuum states of this system (\ref{Vex}) are defined as eigenfunctions with positive total energy $E > 0$ that are delocalized along the $z$-axis and not square-integrable over the full spatial domain. This configuration reveals a spectral structure reminiscent of that described by Gorkov and Dzyaloshinskii \cite{Gorkov} in the strong-field limit, where transverse Landau quantization coexists with free motion along the field direction. Similarly, in the small-radius limit of the cavity, the hard-wall boundary at $\rho = \rho_0$ quantizes the radial ($\rho$) motion into discrete transverse modes with energies $\epsilon_{n_\rho, m}$, while the longitudinal  ($z$) motion remains continuous. The resulting continuum solutions take the form $\Psi(\rho, \phi, z) = R_{n_\rho, m}(\rho)\, e^{i m \phi}\, e^{i k_z z}$, and correspond to band-like continua with energies $E = \epsilon_{n_\rho, m} + k_z^2/2$. This adiabatic separation highlights the deep interplay between external fields and spatial confinement, which governs ionization thresholds, transport phenomena, and the overall spectral structure of Coulomb systems in reduced dimensions.

For small cylinder radii, particularly when $\rho_0 < 1.5$ a.u., the computed energies in Table \ref{tab:param_m0_p0_Bxx} and Figure \ref{parmsBB}(a) are positive, indicating that these states belong to the continuum spectrum rather than the discrete bound-state spectrum. In this regime, the electron is no longer localized and becomes delocalized along the $z$-axis. The solutions correspond to continuum modes characterized by discrete transverse quantization and free longitudinal motion. These continuum states are accurately described within the adiabatic approximation and closely resemble the effective one-dimensional dynamics identified in the strong-field limit by Gorkov and Dzyaloshinskii \cite{Gorkov}, where the transverse confinement—magnetic or geometric—defines the energy thresholds for ionization.

}

\subsubsection{2D \textit{versus} 3D cases}

A remarkable feature of the free hydrogen atom ($B=0,\,\rho_0=\infty$) is 
that the energy of the ground state for the two-dimensional case is 
four times larger than the energy for the three-dimensional case
$E^{(2D)}=4\,E^{(3D)}$. 
In order to establish a similar comparison in the system under discussion, 
it is necessary to consider the two-dimensional hydrogen atom confined within 
a circular region of radius $\rho_0$ in the presence of a magnetic field $B$ 
perpendicular to the plane of the circle. 
In polar coordinates $(\rho,\phi)$, the solution of the Schr\"odinger equation 
can be factorized as $\Psi(\rho,\phi)= (2\pi)^{-1/2}e^{i\,m \,\phi}\,R(\rho)$ 
with $m=0,1,2\dots$. Choosing the vector potential in the symmetric gauge 
${\bf A}=\frac{1}{2}{\bf B}\times \mathbf{\rho}$,
the equation for the radial variable is given by
\begin{equation}
 \left[-\frac{1}{2}\left(\frac{\partial^2}{\partial\,\rho^2}+
 \frac{1}{\rho}\frac{\partial}{\partial\,\rho}\right)
 -\frac{1}{\rho}+\frac{m^2}{2\,\rho^2}+\frac{m}{2}B+\frac{B^2}{8}\rho^2+V_c\right]
 R(\rho)=E\,R(\rho)
\label{hyCB2d}
\end{equation}
where
\begin{equation}
    V_c  =   
\left\{
\begin{array}{ll}
0 & \text{if } \rho < \rho_0\ , \\
\infty & \text{if } \rho \geq \rho_0 \ .
\end{array}
\right.
\end{equation}
The one-dimensional differential equation~(\ref{hyCB2d}) does not admit 
an exact solution; however, it can be solved numerically with the help of 
the MATHEMATICA 13.0 software package using the NDEigensystem command.
When it is possible to compare with the results reported in~\cite{KSSV2020}, 
it results in a complete agreement with the 4 significant digits.

\begin{figure}[!th]
    \includegraphics[scale=1.8]{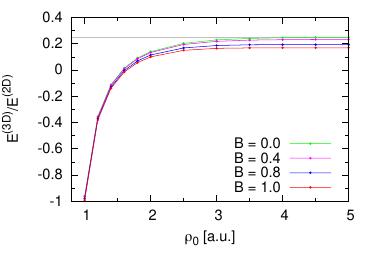}
    \caption{Ratio of the energies $E^{(3D)}/E^{(2D)}$ of the three-dimensional 
    hydrogen atom $E^{(3D)}$ confined by an impenetrable cylindrical cavity 
    of infinite height and the two-dimensional hydrogen atom $E^{(2D)}$ confined 
    to a circular region  as a function of the same cavity radius $\rho_0$
    for the magnetic fields $B=0.0, 0.4, 0.8$ y $1.0$~a.u.}
    \label{fg3d2d}
\end{figure}

In Figure~\ref{fg3d2d}, the ratio of the ground state energies of the three-dimensional 
hydrogen atom confined by an impenetrable infinite cylinder $E^{(3D)}$ and 
the energy of the two-dimensional hydrogen atom confined by an impenetrable 
circular region $E^{(2D)}$ is presented $E^{(3D)}/E^{(2D)}$ as a function 
of the same confinement radius $\rho_0$ for four values of a magnetic field 
$B=0.0, 0.4, 0.8$ and $1.0$~a.u..
For zero magnetic field $B=0.0$, it can be noticed that for confinement 
radii $\rho_0\sim 5.0$~a.u., the ratio approaches the well-known value
$E^{(3D)}/E^{(2D)}=0.25$. As the confining radius $\rho_0$ decreases, 
the energy for both cases $E^{(3D)}$ y $E^{(2D)}$ increases but 
always $|E^{(2D)}|>|E^{(3D)}|$, leading to a gradual decrease in the 
$E^{(3D)}/E^{(2D)}$ ratio. Around $\rho_0\sim 1.57\,$a.u., the sign of the 
energy $E^{(3D)}$ changes from negative to positive {marking the onset of the continuum spectrum}. Consequently, at this critical value of \(\rho_0\), the ratio \(E^{(3D)}/E^{(2D)}\) vanishes. 
For magnetic fields $B\neq 0$ the same qualitative behavior is found
(see Figure~\ref{fg3d2d}).

{
It is important to emphasize a fundamental distinction between the two-dimensional (2D) and three-dimensional (3D) cases: while ionization can occur at small confining radii in 3D systems, such a process is not possible in 2D under the same conditions. This difference arises from a qualitative change in the effective potential and the geometric constraint inherent to two dimensions, where no additional spatial degree of freedom is available for the electron to escape. As a result, the absence of a true continuum in 2D prevents ionization, even when the confining radius is reduced significantly.
}
\subsection{Transversal $\langle \rho \rangle$ and longitudinal size $\langle |z| \rangle$ of the atom}
\label{sizeatom}

A feature of the free hydrogen atom is the fact that when placed in the 
presence of a magnetic field $B$, the magnetic field itself induces a 
cylindrical-type confinement (at large $B$, the electronic cloud of the 
free system develops a needle-like configuration~\cite{TURBINER2006309}).
In order to quantify the deformation of the electron cloud due not only 
to the presence of the magnetic field $B$ but also to the presence of 
the cylindrical cavity, the expectation values 
$\langle\rho\rangle_{\rho_{0}}$ and 
$\langle |z|\rangle_{\rho_{0}}$ are calculated at fixed magnetic field 
$B\in[0.0,1.0]$\,a.u. for different radius of the cylindrical cavity $\rho_0$.

\begin{figure}[!th]
\centering
\subfloat[]{\includegraphics[width=0.45\textwidth]{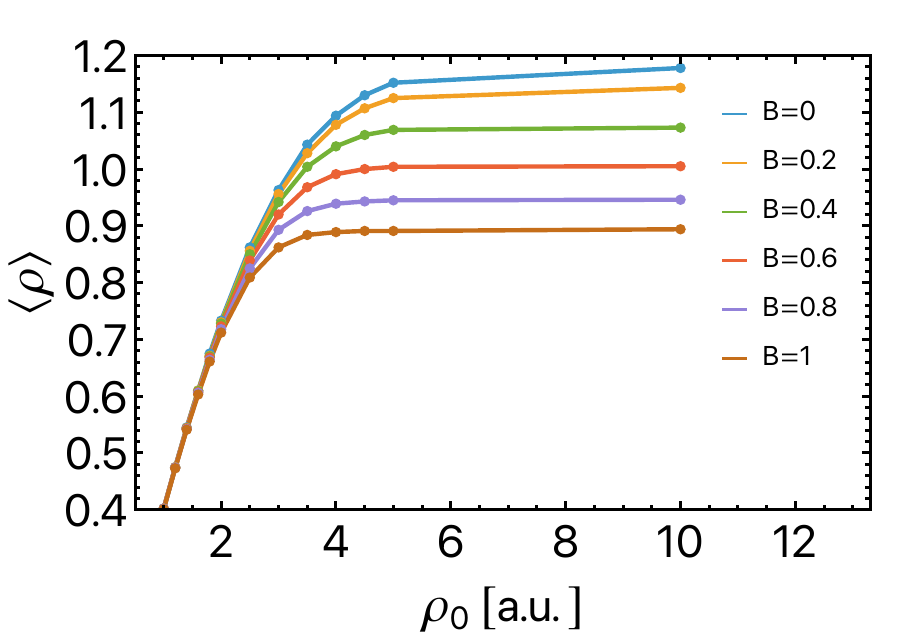}}%
\hfill \hspace{0.2cm}
\subfloat[]{\includegraphics[width=0.5\textwidth]{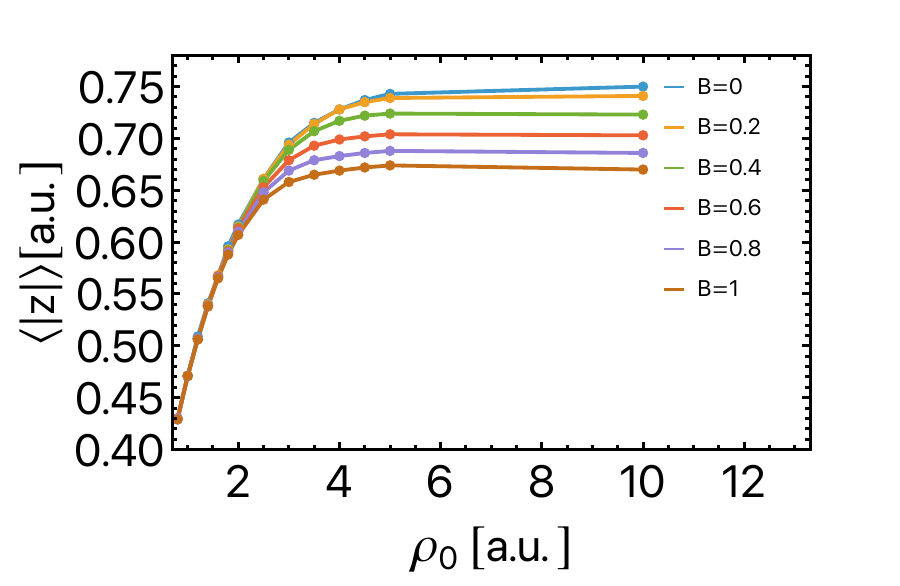}}\\[1ex]

\subfloat[]{\includegraphics[width=0.45\textwidth]{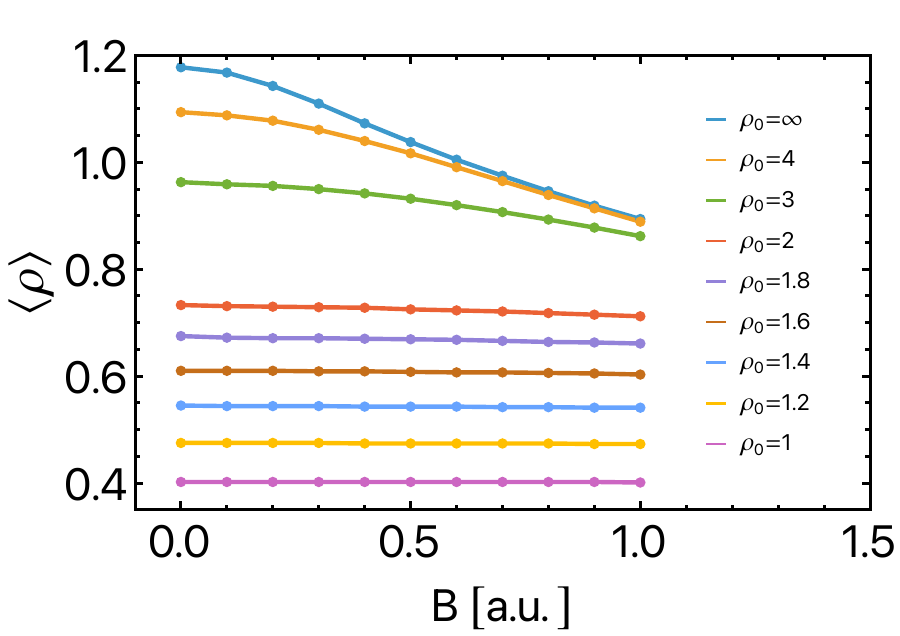}}%
\hfill
\subfloat[]{\includegraphics[width=0.45\textwidth]{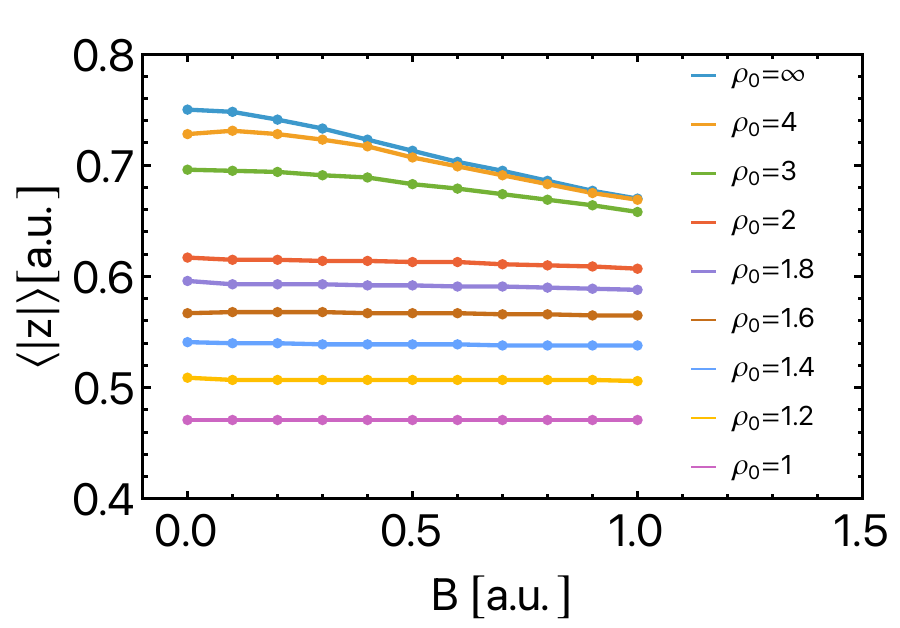}}\\[1ex]

\subfloat[]{\includegraphics[width=0.45\textwidth]{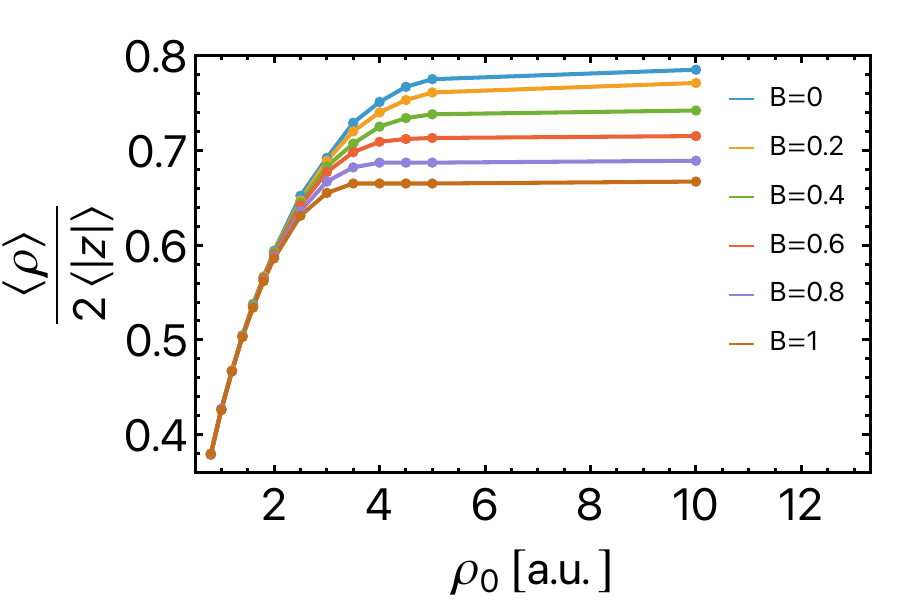}}%
\hfill
\subfloat[]{\includegraphics[width=0.45\textwidth]{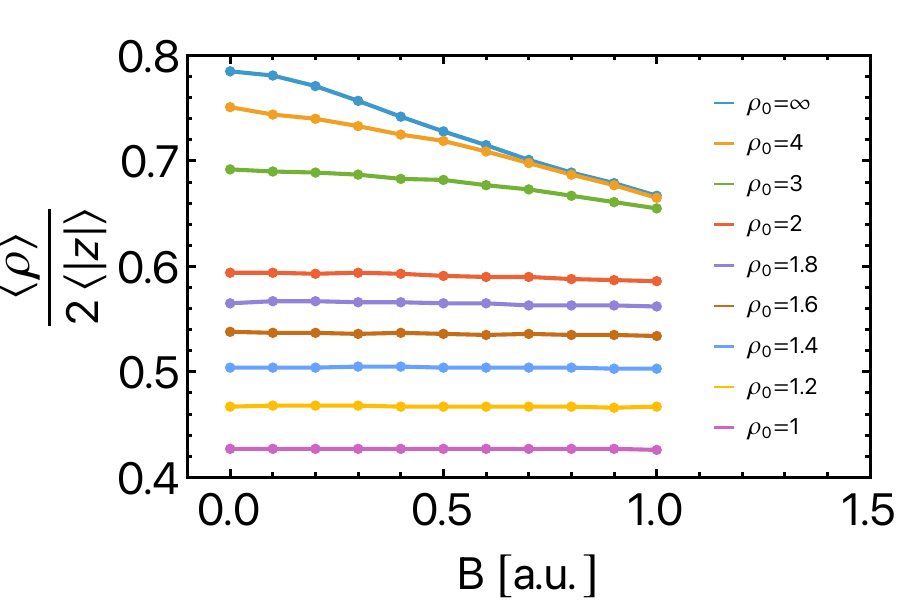}}

\caption{Expectation value of the transversal size (a) $\langle\rho\rangle$ 
and (b) $\langle |z| \rangle$ of the system as a function of the confining 
radius $\rho_0$ for different values of the magnetic fields $B$. 
In (c) and (d) the same expectation values are expressed as a function of the 
magnetic field $B$ for different values of $\rho_0$. The ratio $\frac{\langle \rho \rangle}{2\,\langle |z| \rangle}$ as 
a function of $\rho_0$ and $B$ is depicted in (e) and (f), respectively.}
\label{exvrhoz}
\end{figure}

Table \ref{trho1} presents the dependence of $\langle\rho\rangle_{\rho_{0}}$ 
as a function of the magnetic field for different values of the cavity
radius $\rho_0$ (see also Figure~\ref{exvrhoz} (c)). 
{The calculations of $\langle\rho\rangle_{\rho_{0}}$ for finite values of $\rho_0$, were performed
  with (\ref{psivar}) while for $\rho_0=\infty$, the trial function 
  $\Psi_\infty=(1+\gamma^2\,\rho^2)\,e^{-\alpha\,r-\beta\,B \,\rho^2}$ was used, considering $\gamma$ as an extra variational parameter.}
When the system is not confined ($\rho_0\rightarrow \infty$), it can be seen in 
column 8 that the expectation value $\langle\rho\rangle_{\rho_{0}=\infty}$ 
decreases  with the increasing magnetic field $B$, approaching to the cyclotron 
radius $r_c$ of an electron in a magnetic field $r_c=1/B^{1/2}$. 
For finite values of $\rho_0$ the presence of the cylindrical cavity results in 
a reduction of the expectation value $\langle\rho\rangle_{\rho_{0}}$ relative 
to the free case $\langle\rho\rangle_{\rho_{0}=\infty}$ (see column 7). 
The effect of the magnetic field $B$ on the expectation value 
$\langle\rho\rangle_{\rho_{0}}$ starts to become less important when the 
radius of the cylindrical cavity $\rho_0$ is close to the cyclotron radius 
$\rho_0\sim r_c$ (see column 6). 
Eventually, when the radius $\rho_0$ is smaller than the cyclotron radius
$\rho_0< r_c$, the effect of the magnetic field on 
$\langle\rho\rangle_{\rho_{0}}$ is almost negligible 
(see for instance columns 3, 4 and 5 and Figure~\ref{exvrhoz}\,-(c)).
The dependence of the expectation value $\langle\rho\rangle_{\rho_{0}}$ on 
$\rho_0$ and $B$ is depicted in Figure~\ref{exvrhoz}.
At fixed $\rho_{0}$,  $\langle \rho \rangle_{\rho_{0}}$ is a smooth, decreasing 
function of $B$. However, at fixed magnetic field $B$, it increases as $\rho_0$ grows.

\begin{table}[!th]
  \centering 
  \begin{tabular}{|c|c|c|c|c|c|c|}
    \hline
    $B$  &$B^{-1/2}$ &\ $ \langle  \rho  \rangle _{\rho_{0}=1.8}$  \   & \ $ \langle  \rho  \rangle _{\rho_{0}=2}$  \ &  \ $ \langle  \rho  \rangle _{\rho_{0}=3}$  \ & \ $  \langle  \rho  \rangle _{\rho_{0}=5}$  \ & $ \ \langle \rho \rangle_{\rho_{0}=\infty}$ \ \\
    \hline
  \hline
    0.0  &  & 0.674 & 0.733 & 0.963 & 1.152 & 1.178 \\
    0.1  &3.2&0.672 & 0.731 & 0.959 & 1.143 & 1.168 \\
    0.2  &2.2 & 0.672 & 0.730 & 0.955 & 1.126 & 1.143 \\
    0.4  &1.6&  0.670 & 0.727 & 0.942 & 1.068 & 1.073 \\
    0.6  &1.3&  0.668 & 0.723 & 0.920 & {1.004} & {1.005} \\ 
    0.8  &1.1&  0.665 & 0.718 & 0.893 & {0.945} & {0.946} \\ 
    1.0  &1.0&  0.661 & 0.712 & 0.862 & 0.891 & {0.894} \\    
    \hline
  \end{tabular}
  \caption{Expectation value $\langle \rho \rangle$ as a function 
  of the magnetic field $B$ for different values of the cylindrical 
  cavity radius $\rho_0$. For finite values of $\rho_0$, the results were obtained 
  with (\ref{psivar}) while for $\rho_0=\infty$, the trial function 
  $\Psi_\infty=(1+\gamma^2\,\rho^2)\,e^{-\alpha\,r-\beta\,B \,\rho^2}$ was used
  being $\gamma$ an extra variational parameter.}
  \label{trho1}
\end{table}

Figures~\ref{exvrhoz} (b) and (d) depict the expectation value 
$\langle |z|\rangle$ which provides an estimate of the size of 
the electron cloud in the direction parallel to the magnetic 
field, as a function of the radius $\rho_0$ and the magnetic 
field $B$, respectively.
It is evident that at fixed $B$, $\langle |z|\rangle$ is a smooth, 
bounded increasing function of the radius $\rho_0$, while fixing 
$\rho_0$ it decreases as the magnetic field increases.
As expected, the transversal length $\langle \rho \rangle$ 
is always smaller than the longitudinal size $2\langle |z|\rangle$
The ratio $\frac{\langle \rho \rangle}{2\,\langle |z|\rangle}$ is 
presented in Figures \ref{exvrhoz} (e)-(f) as a function of $\rho_0$
and $B$, respectively.

\section{Shannon entropy in position space }
\label{shanent}

{
The interpretation of probability distributions in quantum mechanics is of fundamental physical relevance, as these distributions provide insight into the position (or momentum) of a system through its wavefunction. Such distributions are characterized by their nodal structure, symmetry constraints, the shape of the density profile (e.g., Gaussian, symmetric, extended), the presence of sharp gradients, and other features reflecting underlying symmetries. Collectively, these attributes define the informational structure of the quantum system.\\
In recent years \cite{Shannon70years}, there has been a growing interest in analyzing quantum systems through the lens of information theory. Within this framework, Shannon entropy has emerged as a particularly valuable tool, as it quantifies the informational content of the probability distribution associated with a given quantum state.
}

Shannon entropy is a key concept in information theory; it was formally introduced by Claude E. Shannon \cite{Shannon} and represents a measure of the uncertainty or unpredictability associated with a set of possible outcomes in a system. In other words, it quantifies the average amount of information expected to be obtained by observing a random event. Specifically, {Shannon entropy quantifies the degree of localization or delocalization of the electron density~\cite{Salazar2021Shannon} and serves as a useful tool for revealing intrinsic structural features of the system.} In this section, the main aim is to analyze the impact of the combined application of two effects, ($i$) an external magnetic field and ($ii$) spatial confinement, on the Shannon entropy in position space.

As we argue in the previous sections, there is a non-trivial competition between the Coulomb potential, the magnetic interaction, and the potential of the hard walls of spatial confinement. It is well known that the physical and chemical properties of confined systems are significantly modified depending on the strength of the confinement, which, in turn, leads to changes in the corresponding entropies.

{ Explicitly, the Shannon entropy $S_{r}$ in position space is given by the following expression:

\begin{equation}
S_{r} = -\int \varrho(\mathbf{r})\, {\rm log}\, \varrho(\mathbf{r}) \,\mathrm{d}^3\mathbf{r} \ ,
\end{equation}
where 
$\varrho(\mathbf{r}) \ = \  | \psi(\rho,r,\varphi)|^{2}$ is the probability density in position space.}

\begin{table}[h]
\begin{center}
\begin{tabular}{*{7}{c|}}
\cline{2-7}
\multirow{2}{*}{ }  & \multicolumn{2}{|c|}{$\rho_{0}=2$ [a.u.]} & \multicolumn{2}{|c|}{$\rho_{0}=5$ [a.u.]} & \multicolumn{2}{|c|}{$\rho_{0}=\infty$} \\
\cline{1-7}
\multicolumn{1}{|c|}{$B$ [a.u.]}  & \multicolumn{1}{c|}{$E$ [a.u.]} & \multicolumn{1}{c|}{$S_{r}$} & \multicolumn{1}{c|}{$E$ [a.u]} & \multicolumn{1}{c|}{$S_{r}$} & \multicolumn{1}{c|}{$E$ [a.u]} & \multicolumn{1}{c|}{$S_{r}$} \\
\hline
\hline
\multicolumn{1}{|c|}{0.0} & -0.277  & 2.941 &   -0.498  & 4.088 &   -0.500  & 4.144 \\
\multicolumn{1}{|c|}{0.1} & -0.276  & 2.935 & -0.496  & 4.072 &   -0.497  & 4.124  \\
\multicolumn{1}{|c|}{0.2} & -0.273  & 2.933 & -0.489  & 4.035 &   -0.490  & 4.071  \\
\multicolumn{1}{|c|}{0.3} & -0.269  & 2.930 & -0.478  & 3.976 &   -0.479  & 3.999   \\
\multicolumn{1}{|c|}{0.4} & -0.263  & 2.925 & -0.464  & 3.905 &   -0.464  & 3.919   \\
\multicolumn{1}{|c|}{0.5} & -0.256  & 2.919 & -0.447  & 3.831 &   -0.447  & 3.837   \\
\multicolumn{1}{|c|}{0.6} & -0.247  & 2.915 & -0.427  & 3.756 &   -0.427  & 3.756   \\
\multicolumn{1}{|c|}{0.7} & -0.236  & 2.906 & -0.405  & 3.680 &   -0.405  & 3.683   \\
\multicolumn{1}{|c|}{0.8} & -0.223  & 2.896 & -0.382  & 3.608 &   -0.382  & 3.610   \\
\multicolumn{1}{|c|}{0.9} & -0.209  & 2.887 & -0.357  & 3.537 &   -0.357  & 3.538   \\
\multicolumn{1}{|c|}{1.0} & -0.194  & 2.876 & -0.330  & 3.472 &   -0.330  & 3.472   \\
\hline
\end{tabular}
\caption{\small {Shannon entropy values in position space for the ground state $m = 0$, $p = 0$ at confinement radii $\rho _{0} = 2.0,\, 5.0$~a.u and without confinement for different magnetic field strength. These values were calculated using the trial function (\ref{psit}).}}
\label{shannon08}
\end{center}
\end{table}

\begin{tabular}{@{}c@{}}
    \begin{minipage}{0.5\textwidth}
         \flushleft      \includegraphics[width=1\linewidth]{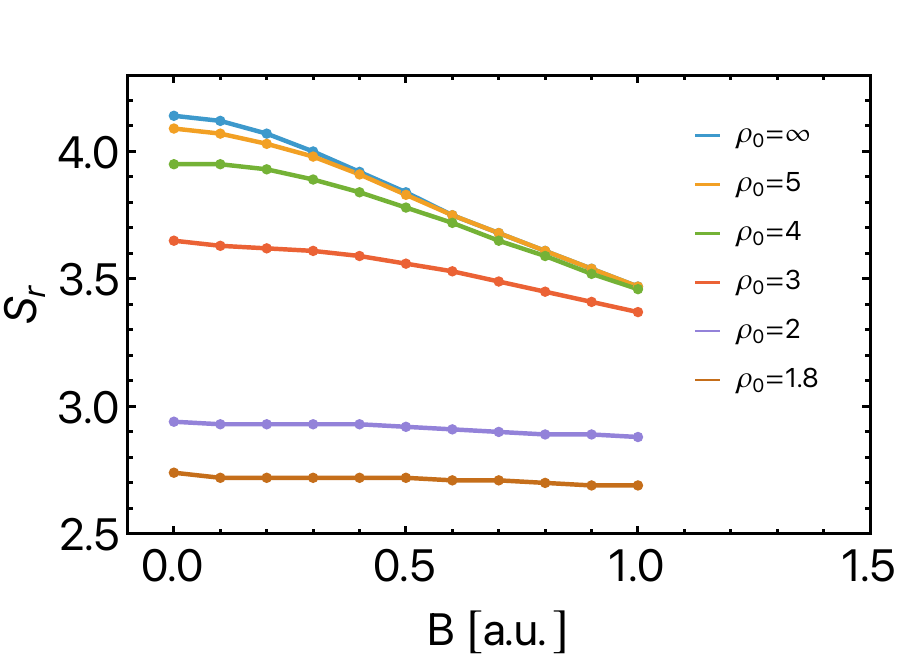}
        \captionof{figure}{{Shannon entropy as a function of the magnetic field $B$.}}
        \label{GShannonB}
    \end{minipage} 
    \hspace{0.02\textwidth}
    \begin{minipage}{0.5\textwidth}
        \flushright
    \includegraphics[width=1\linewidth]{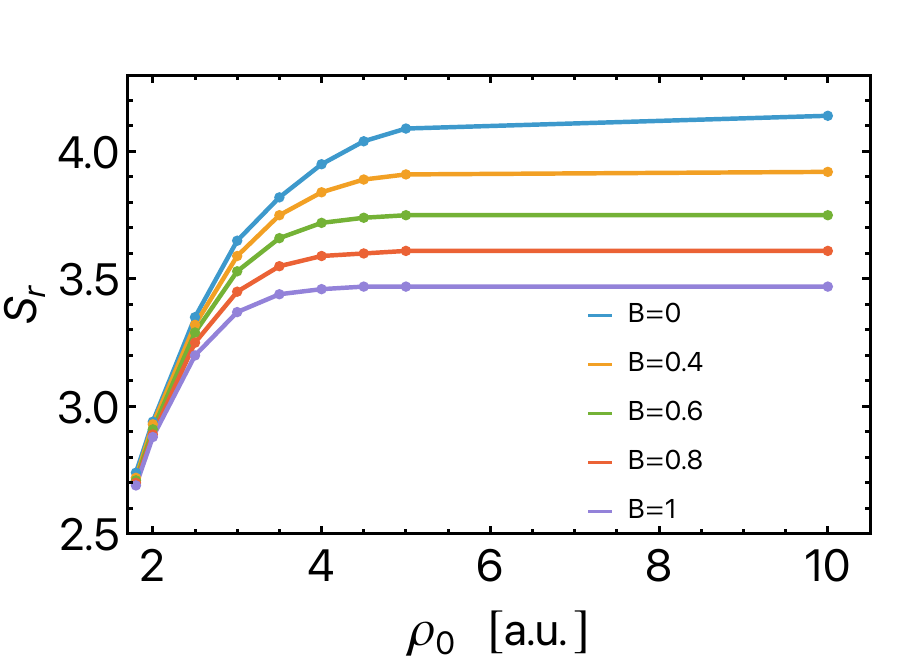}
        \captionof{figure}{{ Shannon entropy as a function of the radius of confinement $\rho_0$. }}
        \label{GShannonrho}
    \end{minipage}
\end{tabular}
\\
{
Shannon entropy quantifies both the amount and the distribution of information encoded in a given probability density, serving as a valuable criterion for evaluating the quality of the wave functions. An accurate wave function should faithfully reflect the expected physical behavior of the system.

In this context, Shannon entropy can be employed as an additional criterion to compare approximate wavefunctions. While two trial functions may yield similar energy values, their corresponding Shannon entropies can differ, revealing discrepancies in their global structure that may not be apparent through energetic or other local observables \cite{Kullback-Leibler}. Therefore, this informational measure provides a complementary perspective for assessing the structural quality or fidelity of an approximate function with respect to a known exact solution.

Shannon entropy is used to evaluate the test function defined in (\ref{psit}), with key results shown in Table \ref{shannon08}. In the weak confinement and zero magnetic field regime, where spherical symmetry is recovered, the entropy matches known values for the free hydrogen atom~\cite{Salazar2021Shannon}, indicating that the test function accurately captures the informational characteristics of the true ground state in this limit.

Figures \ref{GShannonB} and \ref{GShannonrho} illustrate how variations in the magnetic field strength $B$ and the confinement radius induce significant changes in the informational structure of the ground state. Specifically, Figure \ref{GShannonB} shows that, for large values of $\rho_{0}$, increasing the magnetic field leads to a progressive localization of the electronic density. This behavior is further supported by the results in Figure \ref{GShannonrho}, where a reduction in the confinement radius toward the threshold of the discrete spectrum results in a pronounced localization of the system.

These localization trends are consistent with the expected physical behavior, confirming the interplay between magnetic confinement and the spatial constraints imposed by the cylindrical geometry. Consequently, it can be concluded that the trial wave function proposed in this work adequately captures the physical characteristics of the system in the limiting case of the free hydrogen atom (i.e., under weak confinement conditions). 
\\
Consequently, we conclude that the trial function proposed in this work adequately captures the physical characteristics of the system in the limiting case of the free hydrogen atom. This suggests that it constitutes a reliable approximation to the wavefunction describing the quantum states of the confined system under study.}

\section{Conclusions}

A simple compact 3-parametric trial function for the confined 3D hydrogen atom within an impenetrable cylinder of radius $\rho_0$ in a constant magnetic field $B$ is constructed. It codifies the exact symmetries of the system and physically relevant limiting cases. The underlying $(N,m,p)$-representation was clearly explained from both mathematical and physical perspectives. For the whole range of radius values $\rho_0\in [0.8,\,5.0]$ a.u., and magnetic fields $B\in [0.0,1.0]$~a.u., the ground state level ($N=0$, $m=0$, $p=0$) was investigated in detail. A distinctive feature in our trial function is that the cut-off factor $\big(1 - \big(\frac{\rho}{\rho_0}\big)^\nu\big)$,  was determined variationally. It was demonstrated that it results in a notable improvement of the corresponding variational energy. Moreover, compared to more precise purely numerical methods, our results deviate by no more than 1\% with deviations as small as 0.03\%\,.

The energy $E$ and the parameters $(\alpha,\,\beta,\,\nu)$ are always smooth functions of the confinement radius and the magnetic field. The system's localization was systematically quantified using the expectation values 
$\langle \rho \rangle$, $\langle |z| \rangle$, the ratio $\frac{\langle \rho \rangle}{2\langle |z| \rangle}$  as well as the Shannon entropy in position space.

The findings and comparison with existing literature available suggest that this strategy to select the cut-off factor is essential for improving the results. The development of accurate variational functions with strong physical significance plays a crucial role in enhancing our understanding of fundamental phenomena in atomic and molecular systems. The interplay between the magnetic and spatial confinement appears to offer a potential method for measuring the strength of the magnetic field. 

We believe that the present results paves the way to establish a systematic theory (see also \cite{RSR}) on the precise nature of the cut-off factor and its role in the physical description of the confined atomic and molecular systems. Developing systematically improved trial wave functions for confined systems can benefit any field where boundary effects are important, from designing better quantum well lasers to understanding confinement in biological ion channels. For future work, we plan to investigate alternative confining geometries and expand the analysis to include the lowest excited states. Additionally, conducting a similar study for the molecular ion $H_2^+$ would be highly relevant. Extending this variational cut-off approach to two-electron systems like helium could improve our ability to model confined quantum dots with more than one electron, relevant to quantum computing qubits or quantum dot lasers.
In the case of many-electron atoms, the corresponding informational entropic description of the electronic correlation is also an interesting problem. Naturally, the question of finite-mass and relativistic effects remains an intriguing open subject, which will be addressed in future work.

\section*{Credit authorship contribution statement}
\textbf{A. N. Mendoza Tavera} (Validation,  Writing - Original draft preparation, Investigation, Formal analysis) \textbf{H. Olivares-Pilón} (Software, Formal analysis, Methodology, Writing - Original draft preparation, Supervision) \textbf{M. Rodríguez-Arcos} (Validation, Data curation, Supervision) \textbf{A. M. Escobar-Ruiz} (Conceptualization of this study, Formal analysis, Writing - Original draft preparation, Supervision).

\section*{Acknowledgments}

A. M. Escobar Ruiz would like to thank the support from Consejo Nacional de Humanidades, Ciencias y Tecnologías (CONAHCyT) of Mexico under Grant CF-2023-I-1496 and from UAM research grant DAI 2024-CPIR-0. A. N. Mendoza Tavera wishes to thank CONAHCYT (México) for the financial support provided through a doctoral scholarship. A. M. Escobar Ruiz thanks S. A. Cruz for useful personal discussions. M. Rodríguez-Arcos thanks the support from Departamento de Ciencias, EIC, Tecnológico de Monterrey, Campus Ciudad de México. {We would like to thank the anonymous Referees for their insightful remarks and observations, which have significantly contributed to improving the clarity and content of present study.}

\bibliography{Biblio}

\appendix

{
\section{Free unconfined Hydrogen atom: representation $(N,p,m)$ \textit{vs} representation $(n,\ell,m)$}
\label{nmprep}

In this Appendix, for the free 3D hydrogen atom, the explicit relation between the two representations $(N,m,p)$ and $(n,l,m)$ is described. We start by expressing the exact analytical solutions of the 3D hydrogen atom, a quantum (super)integrable central potential 
\[ 
V \ = \ -\frac{k}{r} \ ,
\]
$k>0$, within the representation $(n,\ell,m)$, a standard material of any undergraduate textbook of elementary quantum mechanics  \cite{Landau:1991wop}. In spherical coordinates $(r,\,\theta,\,\varphi)$, assuming an infinitely massive proton located at the origin, the general form of the electronic wave function admits complete separation of variables and reads:

\begin{equation}
\label{solHYsph}
\psi^{(\rm spher)}_{n,\ell,m}(r, \theta, \varphi) \ = \  R_{n,\ell}(r) \,Y_{\ell,m}(\theta,\varphi) \  .
\end{equation}

It depends on three quantized quantum numbers: the principal quantum number $n=1,2, 3 \dots$, the angular momentum $\ell =0,1, \dots, n-1$, and the magnetic quantum number $m= 0, \pm 1, \dots, \pm (\ell-1), \pm \ell $. In (\ref{solHYsph}), $R_{n,\ell}(r)$ describes the radial wavefunction, involving associated Laguerre polynomials and the characteristic exponentially decaying term $e^{-k\,r/n}$, while the spherical harmonic function $Y_{\ell,m}(\theta, \varphi)$ governs the angular part. We call (\ref{solHYsph}) the representation $(n,\ell,m)$.

The corresponding energy spectra, in atomic units ($|e|=1,\,m_e=1,\,\hbar=1$), is given by:
\begin{equation}
\label{EsH}
    E\ =\ -\frac{k^2}{2\,n^2} \ .
\end{equation}
Now, let us take the non-orthogonal coordinate system $(r, \rho, \varphi)$. In these coordinates, it is easy to check that the eigenfunctions of the hydrogen atom can be written in the equivalent form:
\begin{equation}
    \psi_{N,p,m} (r,\rho,\varphi)\ = \ \rho^{|m|} \,z^p \,\chi_{N, p,|m|}\ e^{\sqrt{-2\,E}\,r}\, e^{i\,m\,\varphi} \ .
\label{psiNpm}    
\end{equation}
Here, the (magnetic) quantum number $m$ coincides with the one that appears in the case of spherical coordinates. The quantum numbers $N$ and $p$ refer to those previously defined within the $(N,m,p)-$representation. In (\ref{psiNpm}), the unknown factor $\chi_{N, p,|m|}=\chi_{N, p,|m|}(r,\rho)$ depends solely on the two variables $r$ and $\rho$. It defines the nodal surfaces.

For the parity $\nu=\pm1$ with respect to the reflection $z \rightarrow -z$ we have
\begin{equation}
    \nu\ = \  (-1)^{\ell-|m|} \qquad \rightarrow \qquad \nu \ = \ (-1)^{p} \ ,
\end{equation}
and the quantum number $p=0,1$, in terms of $m$ and $\ell$, reads:
\begin{equation}
    p\ = \ \frac{1}{2}(\,1-(-1)^{\ell-|m|}\, ) \ .
\end{equation}
It implies that all the even values $\ell-|m|=0,2,4,6, \dots$ correspond to $p=0$ (positive parity) whereas the odd values $\ell-|m|=1,3,5,7, \dots$ correspond to $p=1$ (negative parity). Also, 
\begin{equation}
    N\ \equiv \ n-(1+|m|+p) \ ,
    \label{eq:bigN}
\end{equation}
or equivalently, $N\, = \, n_r + \ell - (|m|+p)$ where $n_r$ is the so called radial quantum number.
Hence, in the set of quantum numbers $(N,\,m,\,p)$, the energy spectra (\ref{EsH}) takes the form:
\begin{equation}
    E\ = \ -\frac{k^2}{2\,n^2}\ = \ -\frac{k^2}{2(N + (1+|m|+p))^2}
    \label{eq: energynmp} \ .
\end{equation}
Furthermore, it turns out that the factor $\chi_{N, p,|m|}$ in (\ref{psiNpm}) obeys the spectral problem \cite{Escobar_2023}
\begin{equation}
    h \, \chi(r,\rho) \ = \ k\,\chi(r,\rho) \ ,
\end{equation}
where
\begin{equation}
\label{ho}
\begin{aligned}
h \  = & \ -\,\frac{1}{2}\,r\,\partial^2_r \ - \ \frac{1}{2}\,r\,\partial^2_\rho \ - \ \rho\,\partial^2_{r,\rho}\ - \ \frac{(2 |m| +1)\,r\,-\,2 \, \rho ^2\, \sqrt{-2\,E }}{2\, \rho }\, \partial_\rho
\\ &
-( 1+p+|m| \,-\,r\,\sqrt{-2\,E}  )\,\partial_r \ + \ \sqrt{-2\,E} \,(1+p+|m|)   \ ,
\end{aligned} 
\end{equation}
and the new spectral parameter is $k$. In variables ($r,\,u$), with $u \equiv \rho^2$, the operator $h$ (\ref{ho}) reads
\begin{equation}
\label{halg}
\begin{aligned}
h \  = & \ -\,\frac{1}{2}\,r\,\partial^2_r \ - \ 2\,r\,u\,\partial^2_u \ - \ 2\,u\,\partial^2_{r,u} 
\ - \ 2\,[r\,(1+|m|)\,-\,u\,\sqrt{-2\,E}]\, \partial_u
\\ &
- \, ( 1+p+|m| \,-\,r\,\sqrt{-2\,E}  )\,\partial_r \ + \ \sqrt{-2\,E} \,(1+p+|m|) \ .
\end{aligned}
\end{equation}
Remarkably, the above algebraic operator admits two-variable polynomial eigenfunctions  $\chi_{N, p,|m|}=P_N(r,u)$ of degree $N$ \cite{Escobar_2023}. 
\begin{table}[H]
    \centering \small
    \begin{tabular}{| c | c | c | c | c | c |}
    \hline
     $ n $   &  $\ell$  &  $m$  &  $p$   &  $N$ &  $\chi (r, u)$ \\
     \hline
    \hline
     1       &     0    &   0   &   0    &  0   &     1       \\
     \hline
    
     2       &     0    &   0   &   0    &  1   &     $r-2$       \\
     2       &     1    &  -1   &   0    &  0   &     1             \\
     2       &     1    &   0   &   1    &  0   &     1             \\
     2       &     1    &   1   &   0    &  0   &    $1$            \\
    \hline

     3       &     0    &   0   &   0    &  2   &   $2\,r^2-18\,r+27$       \\
     3       &     1    &  -1   &   0    &  1   &     $r-6$             \\
     3       &     1    &   0   &   1    &  1   &     $r-6$             \\
     3       &     1    &   1   &   0    &  1   &     $r-6$            \\
     3       &     2    &  -2   &   0    &  0   &     1             \\
     3       &     2    &  -1   &   1    &  0   &     1             \\
     3       &     2    &   0   &   0    &  2   &    $2\,r^2-3\,u$            \\
     3       &     2    &   1   &   1    &  0   &     1             \\
     3       &     2    &   2   &   0    &  0   &     1         \\
    \hline
    \end{tabular}
    \caption{\small Representation $(n,\ell,m)\,  $\textit{vs} $(N,m,p)$ for the free hydrogen atom with $k=1$. The eigenpolynomial solutions $\chi (r, u)$ of (\ref{halg}) occurring in $\psi^{(\rm spher)}_{n,\ell,m}(r, \theta, \phi)$ (\ref{solHYsph}) for the $n=1,2,3$ are shown.}
    \label{tab:chi factors}
\end{table}
For the three lowest states $n=1,2,3$, we display in Table \ref{tab:chi factors} the polynomial factor $\chi_{N, p,|m|}$, in variables $r$ and $u=\rho^2$, that occurs in the eigenfunction $\psi^{(\rm spher)}_{n,\ell,m}(r, \theta, \phi)$ (\ref{solHYsph}).

\section{Evolution of the representation $(N, m, p)$ with $B$ and $\rho_0$}
\label{Npmevo}

As revisited in the Appendix \ref{nmprep}, the quantum numbers $(n, \ell, m)$ also characterize the free unconfined hydrogen atom in spherical symmetry. However, $n$ and $\ell$ lose their applicability when a magnetic field or cylindrical confinement is introduced. In such cases, the representation $(N, m, p)$ becomes more appropriate: $m$ remains conserved due to axial symmetry, $p$ reflects parity (also conserved) along the $z$-axis, and $N$ counts nodes in the cylindrical geometry. At zero field and large confinement radius $\rho_0$, both representations are equivalent, but as $B$ increases or $\rho_0$ decreases, the spherical basis breaks down and $(N, m, p)$ captures the physical structure more effectively.

In general, both $m$ and $p$ remain good quantum numbers so long as the system preserves cylindrical symmetry and invariance under $z \to -z$, respectively. The nodal quantum number $N$, by contrast, is not associated with a symmetry of the Hamiltonian, but instead labels the total number of nodal surfaces in the wavefunction, which encodes the transverse and axial structure of the system. The interpretation and spectral role of $N$ evolve significantly depending on the magnetic field strength $B$ and the cylinder radius $\rho_0$.

In the limiting case $B = 0$ and $\rho_0 \to \infty$, the system reduces to the standard hydrogen atom in three-dimensional space. The wavefunction separates into hydrogenic radial and angular components, and $N$ corresponds to the usual nodal structure of the radial wavefunction. Specifically, for fixed $m$ and $p$, the effective principal quantum number is given by $n= N + 1 + |m| + p$, and $N$ counts the radial nodes in $\chi_{N,p,|m|}(\rho, r)$ (\ref{psiNpm}). This regime exhibits the familiar $n^2$-fold degeneracy of the hydrogenic spectrum.

When the cylinder radius $\rho_0$ is finite but the magnetic field remains absent, the hard-wall boundary condition at $\rho = \rho_0$ quantizes the transverse motion. The radial part of the wavefunction resembles a Bessel function with Dirichlet boundary conditions, and $N$ now indexes the number of radial and axial nodes constrained by the boundary. The degeneracy among states with different $|m|$ is lifted due to the centrifugal barrier and confinement.

As the magnetic field is introduced along the $z$-axis, the degeneracy between $\pm m$ is lifted by the Zeeman term, and the diamagnetic term $\propto B^2 \rho^2$ compresses the wavefunction transversely. In the weak to moderate field regime, $N$ continues to label the total number of nodes in $\chi_{N,p,|m|}(\rho, r)$, though the wavefunction shape is modified by magnetic compression. The energy levels are no longer determined solely by $N + |m| + p$, but instead by the interplay of Coulomb attraction, magnetic squeezing, and confinement. 

In the case of a strong magnetic field ($B \gg 1$) and embedded in a cylindrical confinement of large radius $\rho_0$, the electron's transverse motion is dominated by Landau quantization. The magnetic field induces a strong localization in the $x$–$y$ plane, leading to wavefunctions characterized by a Gaussian core modulated by polynomial factors, typical of Landau orbitals. In this regime, the nodal quantum number $N$ naturally separates into two additive components: $N = n_\rho + n_z$, where $n_\rho$ represents the number of radial nodes in the transverse plane (associated with the Landau level index), and $n_z$ counts the number of nodes along the $z$-axis.

Crucially, since the magnetic field does not affect the dynamics along the longitudinal $z$-direction, the axial motion decouples and behaves similarly to a one-dimensional Coulomb problem, possibly modified by axial confinement. The quantum number $n_z$ arises from this separable longitudinal equation and quantifies the number of axial nodes in the wavefunction. If the potential remains symmetric under inversion $z \to -z$, parity along $z$ remains a conserved quantity, and the parity quantum number $p$ is directly related to $n_z$: even $n_z$ corresponds to even parity ($p = 0$), while odd $n_z$ implies odd parity ($p = 1$). In the absence of axial confinement or external perturbations, the system energetically favors the ground longitudinal state with $n_z = 0$ and $p = 0$, corresponding to a symmetric, nodeless profile along $z$. Higher longitudinal excitations with $n_z > 0$ and $p = 1$ (or higher) become accessible only under specific conditions such as axial confinement, field inhomogeneity, or thermal excitation.

Finally, in the limit of strong transverse confinement ($\rho_0 \ll a_0$), the radial motion is severely restricted. Only the lowest transverse mode ($N = 0$) is accessible. The wavefunction becomes quasi-one-dimensional, extended primarily along the $z$-axis. In this regime, the energy spectrum is governed by the longitudinal dynamics, and higher-$N$ states are energetically suppressed due to the confinement and centrifugal barrier.

\begin{table}[h]
\centering
\small 
\caption{Interpretation of the nodal quantum number $N$ in various physical regimes. Here $a_0$ is the Bohr radius.}
\begin{tabular}{@{}l@{\hskip 3.5em}l@{\hskip 3.5em}l@{}}
\toprule
\textbf{Regime} & \textbf{Parameters} & \textbf{Interpretation of $N$} \\ \midrule
Free hydrogen atom & $B = 0$, $\rho_0 \to \infty$ & $N = n - 1 - |m| - p$ \\
Cylindrical confinement & $B = 0$, $\rho_0$ finite & Node index (Bessel zeros) \\
Moderate magnetic field & $B > 0$, $\rho_0$ large & Total nodal surface count \\
Strong magnetic field & $B \gg 1$, $\rho_0$ large & $N = n_\rho + n_z$ \\
Strong confinement & any $B$, $\rho_0 \ll a_0$ & $N = 0$ only (quasi-1D) \\ \bottomrule
\end{tabular}
\end{table}

In summary, the quantum number $N$ provides a useful and flexible labeling of eigenstates in cylindrical geometries. While it retains a clear meaning in the hydrogenic limit, its interpretation becomes more nuanced as magnetic and spatial confinement reshape the underlying wavefunction topology. In the strong field and/or strong confinement regimes, $N$ is effectively replaced by a combination of quantum numbers tailored to the dominant degrees of freedom.

}

\end{document}